\documentclass[iop]{emulateapj}
\usepackage{longtable,lscape}
\usepackage{natbib}
\usepackage[dvips]{color}
\usepackage{multirow}
\usepackage{amsmath}
\usepackage{amsfonts}
\newcommand{\HII}{H\,{\sc ii}}

\newcommand{\OIII}{[O\,{\sc iii}]\,$\lambda$}

\newcommand{\OIIIFIR}{[O\,{\sc iii}]\,88\,$\mu$m}
\newcommand{\OIFIR}{[O\,{\sc i}]\,63\,$\mu$m}
\newcommand{\NII}{[N\,{\sc ii}]}
\newcommand{\CII}{[C\,{\sc ii}]}
\newcommand{\LNII}{$L_{\rm [N\,{\scriptsize \textsc{ii}}]205\,\mu{\rm m}}$}
\newcommand{\LNIIbc}{$L_{\rm [N\,{\scriptsize \textsc{ii}}]122\,\mu{\rm m}}$}
\newcommand{\LNIIbeam}{$L_{\rm [N\,{\scriptsize \textsc{ii}}]205\,\mu m,\,beam}$}
\newcommand{\LIR}{$L_{\rm IR}$}

\newcommand{\NIIIO}{[N\,{\sc ii}]\,205\,$\mu$m}

\newcommand{\fNIIab}{$f_{\rm [N\,{\scriptsize \textsc{ii}}]205\,\mu {\rm m}}$}
\newcommand{\LNIIab}{$L_{\rm [N\,{\scriptsize \textsc{ii}}]205\,\mu {\rm m}}$}
\newcommand{\fNIIbc}{$f_{\rm [N\,{\scriptsize \textsc{ii}}]122\,\mu {\rm m}}$}
\newcommand{\NIIab}{{\rm [N\,{\sc ii}]$\,{205\,\mu {\rm m}}$}}
\newcommand{\NIIbc}{{\rm [N\,{\sc ii}]$\,{122\,\mu {\rm m}}$}}
\newcommand{\fircolor}{$f_{60}/f_{100}$}
\newcommand{\lsolar}{$L_\odot$}
\newcommand{\iso}{{\it ISO}}
\newcommand{\mum}{$\mu$m}
\newcommand{\herschel}{{\it Herschel}}
\newcommand{\iras}{{\it IRAS}}
\newcommand{\kms}{km\,s$^{-1}$}
\newcommand{\LIRPACS}{$L_{\rm IR,\,PACS}$}
\slugcomment{To appear in The Astrophysical Journal}
\begin{document}
%\begin{CJK*}{UTF8}{gbsn}
\shorttitle{The \NII\ 205 $\mu${\MakeLowercase m} Emission in Local LIRGs}
\shortauthors{Zhao et al.}
\title{The \NII\ 205 $\mu${\MakeLowercase m} Emission in Local Luminous Infrared Galaxies$^\star$}
\author{Yinghe~Zhao\altaffilmark{1, 2 ,3}, Nanyao~Lu\altaffilmark{1}, C.~Kevin~Xu\altaffilmark{1}, Yu~Gao\altaffilmark{2, 3}, Steven~D.~Lord\altaffilmark{4}, Vassilis~Charmandaris\altaffilmark{5, 6}, Tanio~Diaz-Santos\altaffilmark{7,8}, Aaron~Evans\altaffilmark{9, 10}, Justin~Howell\altaffilmark{1}, Andreea~O.~Petric\altaffilmark{11}, Paul~P.~van~der~Werf\altaffilmark{12}, and David~B.~Sanders\altaffilmark{13}}
\altaffiltext{1}{Infrared Processing and Analysis Center, California Institute of Technology 100-22, Pasadena, CA 91125, USA; zhaoyinghe@gmail.com}
\altaffiltext{2}{Purple Mountain Observatory, Chinese Academy of Sciences, Nanjing 210008, China}
\altaffiltext{3}{Key Laboratory of Radio Astronomy, Chinese Academy of Sciences, Nanjing 210008, China}
\altaffiltext{4}{The SETI Institute, 189 Bernardo Ave, Suite 100, Mountain View, CA 94043, USA}
\altaffiltext{5}{Department of Physics and ITCP, University of Crete, GR-71003 Heraklion, Greece}
\altaffiltext{6}{IAASARS, National Observatory of Athens, GR-15236, Penteli, Greece}
\altaffiltext{7}{Spitzer Science Center, California Institute of Technology, MS 220-6, Pasadena, CA 91125, USA}
\altaffiltext{8}{Nucleo de Astronomia de la Facultad de Ingenieria, Universidad Diego Portales, Av. Ejercito Libertador 441, Santiago, Chile}
\altaffiltext{9}{Department of Astronomy, University of Virginia, 530 McCormick Road, Charlottesville, VA 22904, USA}
\altaffiltext{10}{National Radio Astronomy Observatory, 520 Edgemont Road, Charlottesville, VA 22903, USA}
\altaffiltext{11}{Gemini Observatory, Northern Operations Center, 670 N. A'ohoku Place, Hilo, HI 96720}
\altaffiltext{12}{Leiden Observatory, Leiden University, PO Box 9513, 2300 RA Leiden, The Netherlands}
\altaffiltext{13}{University of Hawaii, Institute for Astronomy, 2680 Woodlawn Drive, Honolulu, HI 96822, USA}
\altaffiltext{$\star$}{Based on \herschel\ observations. \herschel\ is an ESA space observatory with science instruments provided by European-led Principal Investigator consortia and with important participation from NASA.}

\date{Received:~ Accepted:~}
\begin{abstract}
In this paper, we present the measurements of the \NIIab\ line for a flux-limited sample of 122 (ultra-)luminous infrared galaxies [(U)LIRGs] and 20 additional normal galaxies, obtained with the Herschel Space Observatory (\herschel). We explore the far-infrared (FIR) color dependence of the \NIIab\ (\LNII) to the total infrared  (\LIR) luminosity ratio, and find that \LNII/\LIR\ only depends modestly on the 70-to-160 \mum\ flux density ratio ($f_{70}/f_{160}$) when $f_{70}/f_{160} \lesssim 0.6$, whereas such dependence becomes much steeper for $f_{70}/f_{160}> 0.6$. We also investigate the relation between \LNII\ and star formation rate (SFR), and show that \LNII\ has a nearly linear correlation with SFR, albeit the intercept of such relation varies somewhat with \fircolor, consistent with our previous conclusion that \NIIab\ emission can serve as a SFR indicator with an accuracy of $\sim$0.4 dex, or $\sim$0.2 dex if \fircolor\ is known independently. Furthermore, together with the ISO measurements of \NII, we use a total of $\sim$200 galaxies to derive the local \NIIab\ luminosity function (LF) by tying it to the known IR LF with a bivariate method. As a practical application, we also compute the local SFR volume density ($\dot{\rho}_{\rm SFR}$) using the newly derived SFR calibrator and LF. The resulting $\log\,\dot{\rho}_{\rm SFR} = -1.96\pm0.11$ $M_\odot$\,yr$^{-1}$\,Mpc$^{-3}$ agrees well with previous studies. Finally, we determine the electron densities ($n_e$) of the ionized medium for a subsample of 12 (U)LIRGs with both \NIIab\ and \NIIbc\ data, and find that $n_e$ is in the range of $\sim$$1-100$ cm$^{-3}$, with a median value of 22 cm$^{-3}$.
\end{abstract}

\section{Introduction}
Emission from the forbidden atomic fine-structure transitions in the far-infrared (FIR), such as the [C\,{\sc ii}] 158 \mum, \NII\ 122 \mum\ and 205 \mum, [O\,{\sc i}] 63 \mum\ and 145 \mum, and [O\,{\sc iii}] 52 \mum\ and 88 \mum\ lines, is important for cooling the interstellar medium (ISM), and for providing critical diagnostic tools for the study of the star-forming ISM (e.g. Stacey et al. 1991; Lord et al. 1995; Malhotra et al. 2001; Farrah et al. 2013; De Looze et al. 2014; Fischer et al. 2014; Sargsyan et al. 2014). Among these lines, the [C\,{\sc ii}] 158 \mum\ emission is probably the most important and well studied, since it is the brightest single line in most galaxies and accounts for 0.1-1\% of the total FIR luminosity (e.g. Stacey et al. 1991; Diaz-Santos et al. 2013).

The \NIIIO\ line is of particular interest for the following reasons. Firstly, this $^3$P$_1\,\rightarrow\,^3$P$_0$ transition (205.197 $\mu$m; hereafter \NIIab) of  singly ionized nitrogen is expected to be an excellent indicator of star formation rate (SFR) based on the following facts: 1) the ionization potential of nitrogen is only 14.53 eV. Thus the \NIIab\ emission arises only from \HII\ regions, and essentially traces all warm ionized ISM. It can be utilized to estimate the ionizing photon rate (e.g. Bennett et al. 1994); 2) the \NIIab\ transition can be easily collisionally excited because of its low critical density (44 cm$^{-3}$; Oberst et al. 2006) and excitation energy ($\sim$ 70 K); 3) this emission is usually optically thin and suffers much less dust extinction than optical and near infrared lines. Indeed, Zhao et al. (2013) have shown that the \NIIab\ line can serve as a SFR indicator, which is especially useful for studying high-redshift galaxies for which the redshifted \NIIab\ line is readily obtainable with modern submillimeter telescopes such as the Atacama Large Millimeter/submillimeter Array (ALMA).

Secondly, this line can provide complementary information on the origin of the [C\,{\sc ii}] 158 \mum\ emission (e.g. Oberst et al. 2006; Walter et al. 2009; Decarli et al. 2014; Parkin et al. 2013, 2014; Hughes et al. 2015). The \CII\ line can arise from both neutral and ionized gases since it takes only 11.3 eV to form C$^+$, while the ionization potential of hydrogen is 13.60 eV. Therefore, it is important to know what fraction of the observed \CII\ line emission is from the ionized gas for the study of star-forming regions such as photodissociation region (PDR) modeling. Fortunately, the critical densities for the \NIIab\ and \CII\ 158 $\mu$m lines in ionized regions are nearly identical (44 and 46 cm$^{-3}$ at $T=8000$ K, respectively; Oberst et al. 2006), and both require similar ionization potentials to further form N$^{++}$ (29.6 eV) and C$^{++}$ (24.4 eV). As a result, the \CII/\NIIab\ line ratio from ionized gas is a function of only the assumed N$^+$/C$^+$ abundances within the \HII\ region, and therefore the observed \CII/\NIIab\ line ratio yields the fraction of the \CII\ emission that arises from the ionized gas (Oberst et al. 2006, 2011).

Thirdly, the ratio of the \NII\ 122 \mum\ to 205 \mum\ lines (hereafter $R_{122/205}$) is an excellent density probe of low-density ionized gas due to their different critical densities ($n_{\rm crit}$) required for the collisional excitations and being at the same ionization level. At an electron temperature of 8000 K, $n_{\rm crit}$ are $\sim 293$ and 44 cm$^{-3}$ for the \NII\ 122 and 205 $\mu$m lines, respectively (Oberst et al. 2006). Therefore, $R_{122/205}$ is sensitive to gas densities of $10 \lesssim n_e \lesssim 300$ cm$^{-3}$ (Oberst et al. 2006, 2011; also see \S3.3).

However, the \NIIab\ line is generally inaccessible to ground-based facilities for local galaxies, and for extragalactic objects, only a handful were observed using satellite and airborne platforms (Petuchowski et al. 1994; Lord et al. 1995) prior to the advent of the Herschel Space Observatory (hereafter \herschel; Pilbratt et al. 2010). These studies have shown that the \NIIab\ line is fairly bright, and the luminosity of \NIIab\ line (\LNII) may be up to $\sim$$10^{-3.5}$ times the total infrared luminosity ($L_{\rm IR}$; $8-1000\,\mu{\rm m}$; also see Zhao et al. 2013). With such a high luminosity, this line offers an excellent method for studying SFR and ionized gas properties in galaxies at high redshifts. The advantage of \NIIab\ line over other FIR lines, such as \CII\,158\,\mum, \NIIbc\ and \OIII\,88\,\mum, is that it starts to fall into atmospheric sub/millimeter windows that have higher transmission at lower-$z$, due to its longer wavelength. The detectability  of the \NIIab\ line and its potential for important astrophysical applications at high-$z$ have already been demonstrated by a few experimental ALMA observing campaigns of a galaxy at $z=4.76$ (Nagao et al 2012), and by the IRAM 30m telescope and Plateau de Bure Interferometer detection of distant, strongly lensed galaxies (Combes et al. 2012, $z\sim5.2$; Decarli et al. 2012, 2014, $z\sim3.9$ and 4.7). 

In Zhao et al. (2013), we reported our first results on the \NIIab\ line emission for an initial set of 70 (Ultra-)Luminous infrared galaxies [(U)LIRGs; $L_{\rm IR} \geq 10^{11(12)} L_\odot$]{\footnotemark}{\footnotetext{$L_{\rm IR}$ is calculated by using the {\it IRAS} four-band fluxes and the equation given in Sanders \& Mirabel (1996), i.e., $L_{\rm IR}(8-1000\,\mu{\rm m})=4\pi {\rm D}_{\rm L}^2 f_{\rm IR}$, where D$_{\rm L}$ is the luminosity distance, and $f_{\rm IR}=1.8
\times10^{-14} (13.48 f_{12} + 5.16 f_{25} + 2.58 f_{60} + f_{100})\,({\rm W\,m}^{-2})$}}, observed with the Fourier-transform spectrometer (FTS) of the Spectral and Photometric Imaging Receiver (SPIRE; Griffin et al. 2010) on board {\it Herschel}. In that {\it Letter} we focused on the possibility of using the \NIIab\ emission as a SFR indicator. Here we expand our analysis to our full \herschel\ sample of 122 LIRGs, which is a flux limited  subsample with $f_{\rm IR}(8-1000\, \mu{\rm m}) > 6.5\times10^{-13}$\, W m$^{-2}$ from the Great Observatories All-Sky LIRGs Survey (GOALS; Armus et a. 2009). In addition, we include 20 additional nearby large galaxies for which SPIRE/FTS mapping observations covering the entire galaxy disk are available from the \herschel\ Science Archive (HSA). 

In this paper, besides further investigating the relation between \LNII\ and SFR, we also derive the luminosity function (LF) of the \NIIab\ line and SFR density in the local Universe. It is important to constrain the LF of the \NII\ emission locally since now it becomes possible to build a large sample for studying the \NII\ LF at high-redshift using modern facilities such as ALMA. The local \NII\ LF can serve as a benchmark necessary for observational and theoretical (e.g. Orsi et al. 2014) studies on its evolution. Given the unprecedented sensitivity of \herschel\ at $\sim$200\,\mum, and the large number of galaxies in the local Universe already observed, for the first time we can derive the local \NII\ LF (see \S3.2), using a bivariate method and by utilizing the local IR LF which has been studied extensively in the literature with \iras\ observations (e.g. Soifer et al. 1986; Sanders et al. 2003).

In addition, we estimate the electron densities for a sub-sample of our (U)LIRGs by comparing the observed $R_{122/205}$ with theoretical predications. As shown in Rubin et al. (1994; also see Oberst et al. 2006), $R_{122/205}$ varies from 3 for $n_e\sim100$ cm$^{-3}$ to 10 for $n_e\gtrsim10^3$ cm$^{-3}$. Although Petuchowski et al. (1994) and Lord et al. (1995) observed both the \NIIab\ and \NIIbc\ lines in M82, the use of $R_{122/205}$ for estimating $n_e$ was mostly limited to our own Galaxy (e.g. Wright et al. 1991; Bennett et al. 1994; Oberst et al. 2006, 2011) prior to the advent of \herschel. Furthermore, so far there are only a handful of normal galaxies (e.g. M51 \& Cen A: Parkin et al. 2013, 2014; NGC 891: Hughes et al. 2015) for which $n_e$ of the low-density gas has been derived using the ratio of these two lines. For (U)LIRGs, it is still unclear what is the typical $n_e$ for the low-density ionized gas.

The remainder of this paper is organized as follows: we give a brief introduction of the sample, observations and data reduction in Section 2, present the results and discussion in Section 3, and briefly summarize the main conclusions in the last section. Throughout the paper, we adopt a Hubble constant of $H_0=70~$km~s$^{-1}$~Mpc$^{-1}$, $\Omega_{\rm M} =0.28$, and $\Omega_\Lambda=0.72$, which are based on the five-year WMAP results (Hinshaw et al. 2009), and are the same as those used by the GOALS project (Armus et al. 2009).

\section{Sample, Observations and Data Reduction}
\subsection{(Ultra-)Luminous Infrared Galaxies}
The primary sample studied in this paper is from the \herschel\ open time project {\it Herschel Spectroscopic Survey of Warm Molecular Gas in Local Luminous Infrared Galaxies} (OT1\_nlu\_1; PI: N. Lu). This project aims primarily at studying the dense and warm molecular gas properties of 125 LIRGs (e.g. Lu et al. 2014, 2015), which comprise a flux limited subset of the GOALS sample (Armus et al. 2009). Lu et al. (2015; in preparation) will give the program details and complete set of spectra for individual galaxies. The \NII\ observations were available for 123 targets, one of which is a multiple-source system and our targeted object turned out to be a Galactic source according to its SPIRE/FTS spectrum, consequently excluded from our analysis. Here we present the \NIIab\ data for the 122 galaxies (hereafter GOALS-FTS sample), including 111 LIRGS and 11 ULIRGs. Of these sources, 48 galaxies are point-like sources with respect to the $\sim$17\arcsec\ {\it Herschel} SPIRE/FTS beam at $\sim$210 \mum, and 74 are extended sources, which were determined according to their flux fractions of the FIR continuum emissions at both 70 and 160 \mum\ observed within the SPIRE/FTS beam (see the following subsection for details).

The observations were conducted with the SPIRE/FTS in its point source spectroscopy mode and high spectral resolution configuration, yielding a spectral resolution of 0.04 cm$^{-1}$ (or 1.2 GHz) over the spectral coverage of 194-672 \mum. The data were reduced using the default version of the standard SPIRE reduction and calibration pipeline for point source mode included in the Herschel Interactive Processing Environment (HIPE; Ott 2010) version 11.0. 

In most cases the \NIIab\ line is the brightest line in the SPIRE/FTS wavelength range (Lu et al. 2015, in preparation), and has high signal-to-noise ratios (S/N). As shown in Zhao et al. (2013), the line fluxes were obtained by fitting the observed profile using the instrumental {\it Sinc} function convolved with a free-width Gaussian profile. This is because the line width of most (U)LIRGs is $\gtrsim 200$\,\kms \ (e.g., see Arribas et al. (2015) for ionized gas and Ronsenberg et al. (2015) for molecular gas). Given the instrumental resolution ($\sim$300\,\kms\ at 210\,\mum), the observed line might be marginally resolved and could not be modeled by a pure {\it Sinc} function. Therefore, we adopted the Sinc-convolved-Gaussian (SCG) profiles for the integrated \NII\ line flux measurements except for a few galaxies where a pure {\it Sinc} profile was a better choice due to the intrinsically narrow line width and/or relatively low S/N in the line. During the fitting process, the width of the {\it Sinc} function was fixed to be the SPIRE/FTS resolution (1.2\,GHz), while the width of the Gaussian function was allowed to vary. The resulting full width at half maximum (FWHM) of the \NII\ 205 \mum\ line, which was obtained from the Gaussian part of the SCG profile, is $\sim 100 - 600$ km s$^{-1}$, with a median value of $\sim 300$ km s$^{-1}$. Based on the $1\sigma$ statistical uncertainties, the lines in most ($>80\%$) sources are detected at better than 7$\sigma$, with the median at $\sim$14$\sigma$. The measured line fluxes are given in Table \ref{fluxes}.

\begin{deluxetable*}{lcccccccc}
%\tablenum{2}
\centering
\tablecaption{Fluxes of the \NIIab\ Emission{\label{fluxes}}}
\tablewidth{\textwidth}
%\tablewidth{86mm}
\tabletypesize{\scriptsize}
\tablehead{
\colhead{Galaxy}&\colhead{R.A.}&\colhead{Decl.}&\colhead{Obs ID}&\colhead{$f_{{\rm [N\,{\scriptsize \textsc{ii}}]}\,205\mu{\rm m}}${\tablenotemark{a}}}&\colhead{$f_{\rm corr}$}&\colhead{SFR}&\colhead{$f_{60}$}&\colhead{$f_{100}$}\\
\colhead{Name}&\colhead{(hh:mm:ss)}&\colhead{(dd:mm:ss)}&\colhead{}&\colhead{(10$^{-17}$ W m$^{-2}$)}&($R_{122/205})$&($M_\odot\,{\rm yr}^{-1}$)&(Jy)&(Jy)\\
\colhead{(1)}&\colhead{(2)}&\colhead{(3)}&\colhead{(4)}&\colhead{(5)}&\colhead{(6)}&\colhead{(7)}&\colhead{(8)}&{(9)}
}
\startdata
\multicolumn{9}{c}{GOALS-FTS Sample}\\
\hline\noalign{\smallskip}
NGC0023  & $00:09:53.4$ & $+25:55:26.2$ & 1342247622 &$10.79\pm0.75$ &  1.58  &23.0 &  9.03 & 15.66\\
NGC0034  & $00:11:06.5$ &$-12:06:24.9$ & 1342199253 &$2.55\pm 0.49$&   2.33  &55.0  &17.05  &16.86\\
MCG-02-01-051 &$00:18:50.9$ &$-10:22:37.6$ &1342247617 & $3.80\pm0.54$   &1.00 & 50.3 &  7.48 &  9.66\\
\nodata &\nodata & \nodata & \nodata & \nodata & \nodata &\nodata &\nodata &\nodata\\
\hline\noalign{\smallskip}
\multicolumn{9}{c}{Mapping Galaxy Sample}\\
\hline\noalign{\smallskip}
NGC1266&$03:16:00.7$&$-02:25:38$&1342239353&$  2.85\pm  0.42$&--- &  5.6&  13.13&  16.89\\
NGC1377&$03:36:39.1$&$-20:54:08$& 1342239352&$  0.58\pm  0.10$&--- &\nodata &   7.43&   5.95\\
NGC1482&$03:54:38.9$&$-20:30:10$& 1342248233&$ 36.1\pm  2.4$&--- & 13.1&  33.96&  46.73\\
\nodata &\nodata & \nodata & \nodata & \nodata & \nodata &\nodata &\nodata &\nodata\\
\hline\noalign{\smallskip}
\multicolumn{9}{c}{ISO Sample}\\
\hline\noalign{\smallskip}
NGC0520&$01:24:34.90$&$+03:47:30.0$& 77702295&$  39.2\pm  16.3$&1.2& 15.4& 31.10& 47.12\\
NGC0986&$02:33:34.10$&$-39:02:41.0$& 74300187&$  43.3\pm  18.1$&1.2& 12.6& 25.14& 51.31\\
NGC1222&$03:08:56.80$&$-02:57:18.0$& 82400836&$   8.7\pm   4.0$&1.5&  8.3& 13.07& 15.38\\
\nodata &\nodata & \nodata & \nodata & \nodata & \nodata &\nodata &\nodata &\nodata\\
\enddata
\tablecomments{Columns: (1) galaxy name; (2) and (3) right ascension and declination (J2000); for the GOALS-FTS sample,  the coordinate gives the position where the \herschel\ SPIRE/FTS observation was pointed; (4) observation ID (number) for the \herschel\ ({\it ISO}) observation; (5) \NIIab\ flux: measured from the SPIRE/FTS spectra for the GOALS-FTS and Mapping Galaxy samples; obtained from \NIIbc\ emission for the ISO sample; (6) for the GOALS-FTS sample, correction factor ($f_{\rm corr}$) applied to Column (5) to obtain the total \NIIab\ flux (see \S2.1.1 for details); for the ISO sample, the \NIIbc-to-\NIIab\ conversion factor ($R_{21})$; (7) star formation rate (\S3.1); (8) and (9) \iras\ fluxes at 60 and 100 \mum\ respectively.}
\tablenotetext{a}{For the ISO sample, the listed error has taken into account for the uncertainty of $R_{122/205}$. \\ \vspace{5mm} (This table is available in its entirety in a machine-readable form in the online journal. A portion is shown here for guidance regarding its form and content.)}
\end{deluxetable*}

\subsection{Local Normal Galaxies}
\begin{figure}[tbp]
\centering
\includegraphics[width=0.48\textwidth,bb = 4 8 408 462]{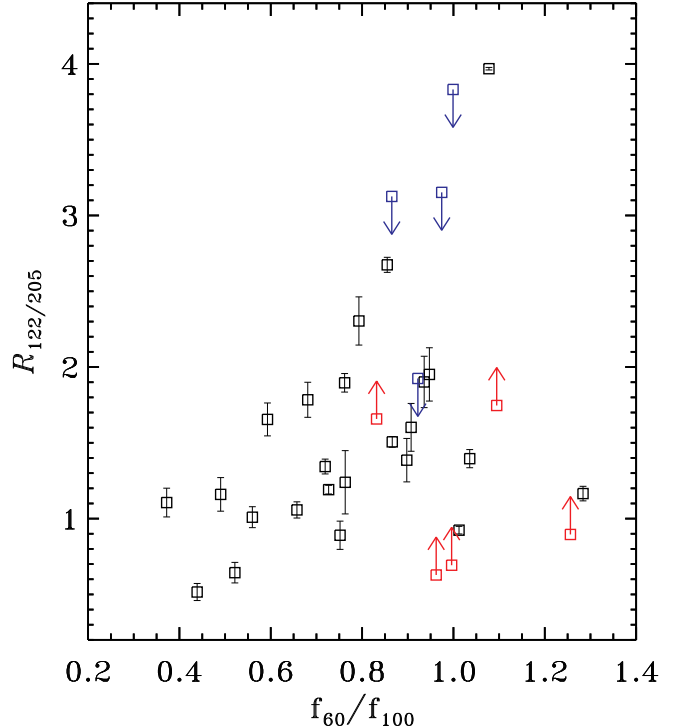}
\caption{The \NIIbc\ to \NIIab\ emission ratio ($R_{122/205}$) plotted against the {\it IRAS} FIR color. The upward and downward arrows represent lower and upper limits, respectively.}
\label{Figniiratio}
\end{figure}

Almost all of our sample galaxies are (U)LIRGs, and hence have a rather limited dynamic range of several physical parameters such as luminosity, FIR color, etc. To increase the sample size and dynamic range of our study, we also include in our analysis 20 nearby normal galaxies having \herschel\ SPIRE/FTS mapping observations that cover the {\it entire} galaxy. As in Zhao et al. (2013), we further include another 53 unresolved galaxies (23 detections and 30 upper limits) observed by the {\it Infrared Space Observatory} (\iso; Kessler et al. 1996, 2003) Long Wavelength Spectrometer (LWS), for which the \NIIbc\ fluxes were measured by Brauher et al. (2008; hereafter \iso\ sample).

\subsubsection{\herschel\ SPIRE/FTS Mapping Observations}
These observations were carried out by various \herschel\ projects [e.g. {\it Very Nearby Galaxies Survey} (KPGT\_cwilso01\_1; PI: C. D. Wilson; e.g. Spinoglio et al 2012b; Parkin et al. 2013; Schirm et al. 2014; Hughes et al. 2015) and the {\it Beyond the Peak: Resolved Far-Infrared Spectral Mapping of Nearby Galaxies with SPIRE/FTS} (OT1\_jsmith01\_1; PI: J. D. Smith)]. The level 0.5 raw data were obtained through HSA, and reduced with the default pipeline for mapping observations provided in HIPE 11. We then added up all pixels of the level 2 product which have valid data, and measured the integrated \NIIab\ fluxes using the method described in Section 2.1 above. During this process, we have converted the units of the mapping data from MJy/sr to Jy/pixel using the area of an individual pixel.

\subsubsection{\iso\ LWS Observations}
For the \iso\ sample, we derived their \NIIab\ fluxes (\fNIIab) from the observed \NIIbc\ fluxes (\fNIIbc) using the following empirical method. The $R_{122/205}$ used in our conversion was estimated on the basis of the actual observations of these two lines for a sample consisting of 7 normal galaxies and 26 (U)LIRGs. Besides our GOALS-FTS sources, about a half of these (U)LIRGs are from Farrah et al. (2013) sample, for which the \NIIbc\ were observed with \herschel\ PACS, and \fNIIbc\ were adopted from Farrah et al. (2013); whereas \fNIIab\ were measured from the SPIRE/FTS data observed in the program ``OT1\_dfarrah\_1" (PI: D. Farrah).  For our GOALS-FTS objects, the aperture-corrected (see \S2.3) \fNIIab\ were used.

 In Figure \ref{Figniiratio} we plotted $R_{122/205}$ against the {\it IRAS} FIR color, \fircolor. It seems that $R_{122/205}$ shows some dependence on \fircolor. Kewley et al. (2000) found that electron density tends to correlate with $f_{25}/f_{60}$. Therefore, the weak dependence of $R_{122/205}$ on \fircolor\ appears to be understandable. To further check whether there is a correlation between \fircolor\ and $R_{122/205}$, we computed the Kendall's $\tau$ correlation coefficient using the {\it cenken} function in the {\it NADA} package within the public domain {\bf R} statistical software environment{\footnotemark}{\footnotetext{\url{http://www.R-project.org/}}}. For the whole dataset presented in Figure \ref{Figniiratio}, we have $\tau=0.18$, with a p-value of 0.13, and thus we do not reject the null hypothesis that these two parameters are uncorrelated at the 0.05 significance level. However, we have $\tau=0.35$ with a p-value of 0.03 if we limit the sample to $f_{60}/f_{100} < 0.9$. Therefore, there exists a weak correlation between \fircolor\ and $R_{122/205}$ within this color range. Since almost all of the {\it ISO} galaxies fall within this color range, we adopt FIR color-dependent $R_{122/205}$. Nevertheless, we caution that such an analysis is possibly limited by the small size of the sample. However, this (un)correlation will not affect our main conclusions since the two $R_{122/205}$ values adopted in the following only differ by $\sim$0.3, which is negligible compared to the overall uncertainties. 

For sources with $f_{60}/f_{100} < 0.7$, we adopt $R_{122/205} = 1.2\pm0.5$, and for the other warmer galaxies (with $f_{60}/f_{100} <1.0$), $R_{122/205}=1.5\pm0.7$. These adopted $R_{122/205}$ values are the median of the corresponding detections, and are lower than the single value of 2.6 adopted in Zhao et al. (2013). The latter was based on the theoretical prediction for an electron density of $n_e=80$ cm$^{-3}$, i.e. the median value of \HII\ regions in late type galaxies (Ho et al. 1997). However, the adopted $n_e$ in Zhao et al. (2013) was measured from \HII\ regions in the centers of nearby galaxies, and thus might be an overestimate of the mean $n_e$ for the entire galaxy. As a result, the overall \fNIIab\ obtained from \fNIIbc\ in Zhao et al. (2013) was somewhat underestimated.

For these nearby galaxies, the redshift-independent distance was adopted if a direct primary measurement distance could be found in the NED\footnote{http://ned.ipac.caltech.edu} database, otherwise it was derived with the same method as used for our (U)LIRG sample (e.g. Armus et al. 2009), i.e. by correcting the heliocentric velocity for the 3-attractor flow model of Mould et al. (2000). 

\subsection{Aperture Corrections}
Around 205\,\mum\, the SPIRE/FTS beam can be well represented by a symmetrical Gaussian profile with a FWHM of 17\arcsec\ (Makiwa et al. 2013; Swinyard et al. 2014). However, this beam can not fully cover the entire \NII\ emission region in most of our targets assuming their FIR sizes indicate the extent of the \NII\ emission. To define a source as ``extended" compared to the SPIRE/FTS beam, we calculated the fractional 70 and 160\,\mum\ fluxes within a Gaussian beam of FWHM of 17\arcsec\ (see below). An extended source will have both fractions less than 90\%. Based on this definition, 74 galaxies are classified as extended. Therefore, to achieve our ultimate goals of deriving the LF of the \NIIab\ emission, as well as of further exploring the applicability of \LNII\ as a SFR indicator, we need to apply an aperture correction to the observed \NIIab\ fluxes for most sources. 

Zhao et al. (2013) found that \LNII\ correlates almost linearly with \LIR. Hence, the aperture correction can be done by utilizing PACS photometry images and estimating the \LIR\ measured within the region outside the SPIRE/FTS beam. However, as already shown in Zhao et al. (2013), the \LNII/\LIR\ ratio also depends somewhat on the FIR color. To account for this dependence and to minimize the uncertainty in the final, total \LNII, we used the FIR color-dependent \LNII-\LIR\ relation (see below) to correct the \LNII\ which was measured directly from the SPIRE/FTS spectra.

To measure the \LIR\ and FIR color within the 17\arcsec\ SPIRE/FTS beam near 205 \mum\ for our sample of (U)LIRGs, we applied the following steps. Firstly, in order to have the same resolution as the SPIRE/FTS, we convolved the 70 and 160 \mum\ images (e.g. Chu et al. 2015, in preparation), which were obtained with the Photodetector Array Camera and Spectrometer (PACS; Poglitsch et al. 2010), with kernels computed through the algorithm described in Aniano et al. (2011). These convolution kernels were generated by comparing the PACS PSFs at 70\,\mum\ and 160\,\mum\ with a Gaussian profile of FWHM of 17\arcsec\ (as a representative of the SPIRE/FTS beam around 210\,\mum). Then we converted the units of the convolved images from Jy/arcsec$^2$ to Jy/beam by multiplying the area of the Gaussian profile. The fluxes within the SPIRE/FTS beam at 70 and 160 \mum\ (hereafter $f_{70,\,{\rm beam}}$ and $f_{160,\,{\rm beam}}$, respectively) were measured from the convolved PACS images at the SPIRE/FTS pointing position. The total fluxes ($f_{70,\,{\rm tot}}$ and $f_{160,\,{\rm tot}}$), were also measured from the convolved images by doing aperture photometry. Therefore, the fluxes outside the SPIRE/FTS beam are $f_{70,\,{\rm out}} = f_{70,\,{\rm tot}}- f_{70,\,{\rm beam}}$ and $f_{160,\,{\rm out}} = f_{160,\,{\rm tot}}- f_{160,\,{\rm beam}}$, for the 70 and 160\,\mum\ respectively. Note that in this subsection the {\it IR luminosity ($L_{\rm IR,\,PACS}$)} is calculated using the $f_{70}$ and $f_{160}$ fluxes and the formula ($L_{\rm IR,\,PACS}=1.010\nu L(70\,\mu{\rm m})+1.218 \nu L(160\,\mu {\rm m})$) presented in Galametz et al. (2013) since the PACS data have much higher angular resolution than the \iras\ data, which is necessary for our purpose. However, the \LIR\ used for the remainder of our analysis is derived from the {\it IRAS} four-band fluxes and the well known equation given in Sanders \& Mirabel (1996). 

\begin{figure*}[tbp]
\centering
\includegraphics[width=0.8\textwidth,bb = 10 125 572 440]{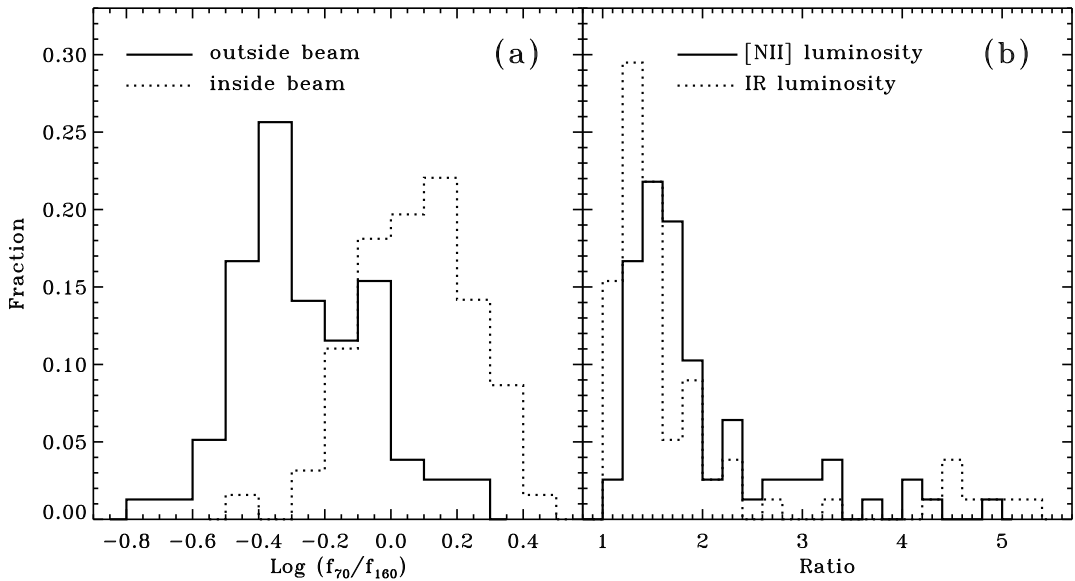}
\caption{Distributions of (a) the FIR colors measured inside (dotted line) and outside (solid line) the SPIRE/FTS beam. (b) the ratios of aperture-corrected \LNII\ to \LNIIbeam\ (solid) and the total $L_{\rm IR,\,PACS}$ to $L_{\rm IR,\,PACS,\,beam}$ (dotted), for the subsample of our extended (U)LIRGs}
\label{Figdist}
\end{figure*}

Since the galaxies in our sample are (U)LIRGs, and our SPIRE observations usually were targeted at the center of each object, the measured FIR color within the beam is very warm, whereas the part missed by the SPIRE beam, which needs to be corrected for, is generally much colder. This is illustrated in Figure \ref{Figdist}a, in which we plotted the distributions of the FIR color inside ($(f_{70}/f_{160})_{\rm beam}$; dotted histogram) and outside ($(f_{70}/f_{160})_{\rm out}$; solid histogram) the SPIRE/FTS beam. We can see that $(f_{70}/f_{160})_{\rm beam}$ and $(f_{70}/f_{160})_{\rm out}$ peak at $\sim$1.41 and $\sim$0.45 respectively.  Therefore, it is necessary to include more fiduciary data with cooler FIR colors to better establish the \LNII/\LIR-FIR color relationship. 

For this purpose, we include in our analysis a dozen nearby, spatially resolved galaxies, which have SPIRE/FTS mapping observations in the HSA and are mainly from the same projects listed in section 2.2.1. These SPIRE/FTS observations (3 of them are in the sample of the 20 galaxies mentioned in \S2.2) were reduced with the same method as described in \S2.2.1. The PACS imaging data of these nearby, very extended galaxies were reduced using the Scanamorphos technique (Roussel 2013) provided in HIPE 12.1, and then were convolved from their native resolutions to the 17\arcsec\ resolution of SPIRE at $\sim$210\,\mum\ using the same method as for our GOALS-FTS sample. The convolved images were rebinned to maps with pixel size corresponding to the SPIRE/FTS mapping observations. In order to increase the S/N for the SPIRE/FTS mapping observations, we stacked spectra from regions having similar $f_{70}/f_{160}$ colors. Also, to reduce the uncertainties in the stacked spectrum and IR flux for each color bin, only pixels with ${\rm S/N} > 3$ both at 70 and 160\,\mum\ were used. The \NIIab\ fluxes of the stacked spectrum from these galaxies were measured using the same method as for the GOALS-FTS sample described in \S2.1 above. 

\begin{figure*}[tbp]
\centering
\includegraphics[width=0.7\textwidth,bb = 36 8 488 488]{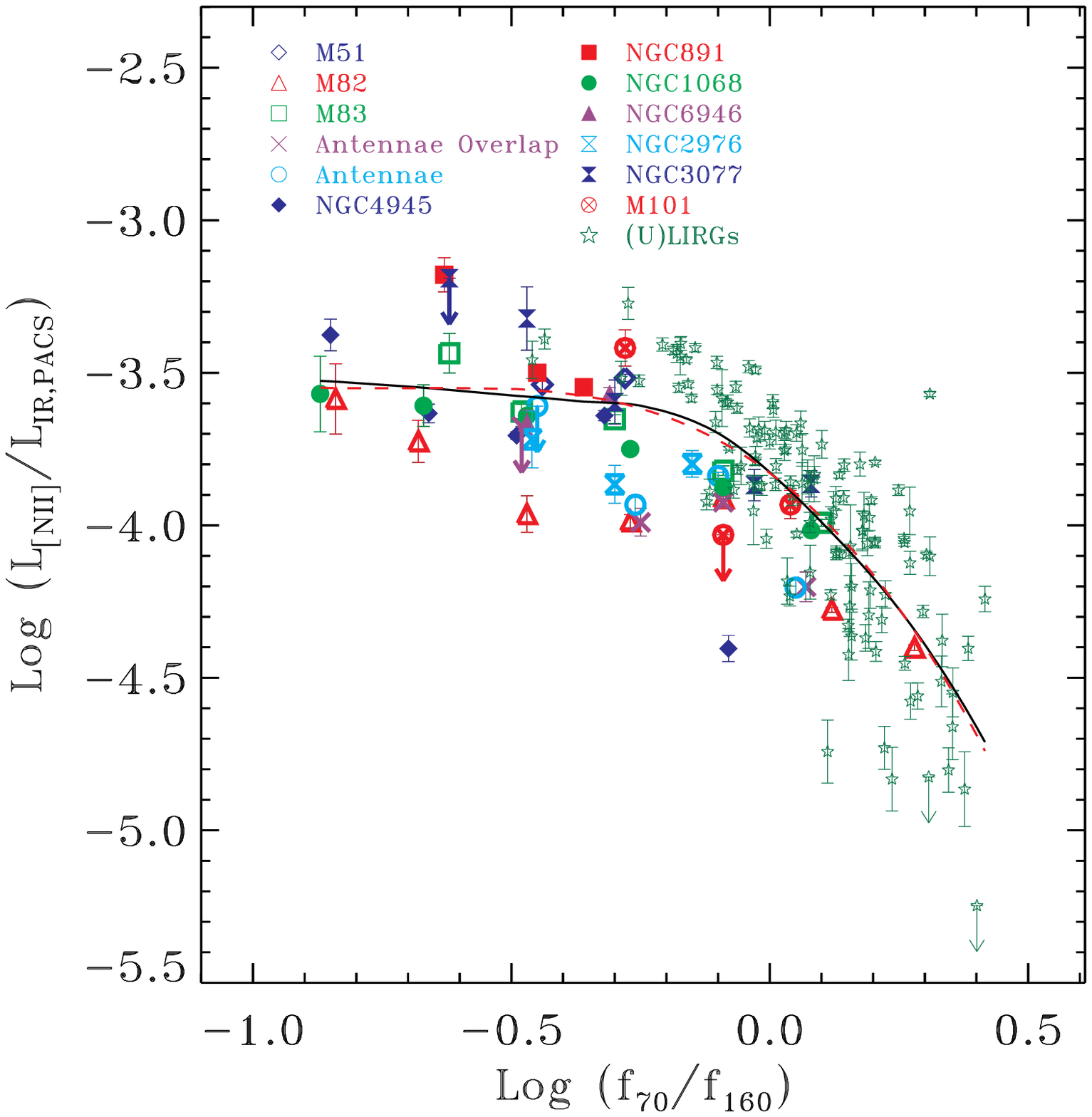}
\caption{Correlation between the \NIIab\ to IR luminosity (see the text for the derivation of the IR luminosity used here) ratio and FIR color. For (U)LIRGs, \LNII, $L_{\rm IR,\,PACS}$ and FIR color were measured within the SPIRE beam, whereas for other labelled individual galaxies, we used the SPIRE mapping observations and stacked spectra for similar FIR colors to measure \LNII\ and $L_{\rm IR,\,PACS}$ within the region of the mapped pixel (see the text for more details). The dashed (red) line shows the best polynomial fit to the results shown by the solid (black) line, which were computed using the {\it locfit.censor} function.}
\label{Figcolordep}
\end{figure*}

The final \LNII/\LIRPACS-FIR color relation is shown in Figure \ref{Figcolordep}, where \LNII, $L_{\rm IR,\,PACS}$ and FIR color were measured within the SPIRE/FTS beam for all of our GOALS-FTS (U)LIRGs, and for other galaxies these were measured within the stacked spaxels. This relation is rather flat for $\log (f_{70}/f_{160})\lesssim -0.2$ (equivalent to $f_{60}/f_{100} \lesssim 0.46$; after Dale et al. 2001), but becomes much steeper when the FIR color is getting warmer, and has the largest scatter at the warmest end. To investigate this relation, we used the Kaplan-Meier estimate (Kaplan \& Meier 1958) for censored data{\footnotemark}{\footnotetext{Implemented in the {\it locfit.censor} function in the {\it locfit} package in \bf{R}}}. The resulting \LNII/$L_{\rm IR,\,PACS}$-$f_{70}/f_{160}$ relation is shown by the solid line in Figure \ref{Figcolordep}, with a scatter of 0.22 dex (compared to the observed \LNII/$L_{\rm IR,\,PACS}$). We used this relation to calculate the \LNII\ outside the SPIRE/FTS beam for our GOALS-FTS sample, and the uncertainty of 0.22 dex was propagated to the final uncertainty values for \LNII\ after taking a quadratic sum of all errors.

We also fitted the relation estimated by {\it locfit.censor} (i.e. the solid line in Figure \ref{Figcolordep}) with  a third-order polynomial function because  an analytical form could be more convenient for future studies. The best-fit gives
\begin{equation}
\log\,(L_{\rm [N\,{\scriptsize \textsc{ii}}]205\,\mu m}/L_{\rm IR}) = -3.83 - 1.26x - 1.86x^2 -0.90x^3,
\end{equation}
with $x = \log\, (f_{70}/f_{160})$, and a scatter of 0.01 dex, and it is plotted as a dash line in Figure \ref{Figcolordep}. Please note this relation is only valid for the color range we have investigated, i.e. $-0.9\leq\log\, (f_{70}/f_{160})\leq 0.4$. An uncertainty of 0.23 dex should be adopted if this best-fit relation is used to compute \LNII/\LIR\ from $f_{70}/f_{160}$. As seen in Figure 3, there is a strong relation between \LNII/\LIR\ and FIR color for $\log\,(f_{70}/f_{160})_{\rm out}>-0.2$. Given that about one third of our extended sources have $\log\,(f_{70}/f_{160})_{\rm out}>-0.2$ (see Figure \ref{Figdist}a) it is necessary to use a color-dependent aperture correction for the \NIIab\ emission.

The aperture-corrected, total \LNII\ for the extended GOALS-FTS sources were, $L_{\rm [N\,{\scriptsize \textsc{ii}}]205\,\mu m} = L_{\rm [N\,{\scriptsize \textsc{ii}}]205\,\mu m,\,beam}+L_{\rm [N\,{\scriptsize \textsc{ii}}]205\,\mu m,\,out}$, where $L_{\rm [N\,{\scriptsize \textsc{ii}}]205\,\mu m,\,beam}$ and $L_{\rm [N\,{\scriptsize \textsc{ii}}]205\,\mu m,\,out}$ represent the \NIIab\ luminosities measured inside and outside the SPIRE/FTS beam, respectively. $L_{\rm [N\,{\scriptsize \textsc{ii}}]205\,\mu m,\,beam}$ was measured directly from the SPIRE/FTS spectrum, while $L_{\rm [N\,{\scriptsize \textsc{ii}}]205\,\mu m,\,out}$ was obtained using $L_{\rm IR,\,PACS,\,out}$ and $(f_{70}/f_{160})_{\rm out}$. As shown in Figure \ref{Figdist}a, most of our galaxies have $\log\,(f_{70}/f_{160})_{\rm out}<-0.2$, so the aperture correction for these sources should not be very sensitive to the FIR color as indicated by Figure \ref{Figcolordep} (and equation 1). The solid histogram in Figure \ref{Figdist}(b) shows the distribution of the \LNII/\LNIIbeam\ ratios ($\equiv f_{\rm corr}$) for the extended sources. For about 70\% of all the cases, $f_{\rm corr}$ is less than 2, i.e. the SPIRE/FTS beam captured more than a half of the total \NIIab\ emission from a galaxy. 

To further examine whether the color-dependent aperture correction is essential, we also plotted the \LIRPACS/$L_{\rm IR,\,PACS,\,beam}$ distribution (dotted line) in Figure \ref{Figdist}b. The median values of $f_{\rm corr}$ and \LIRPACS/$L_{\rm IR,\,PACS,\,beam}$ are 1.66 and 1.44, respectively, which indicates that the overall \LNII\ would be underestimated by about 12\% if we used a constant \LNII/\LIR\ ratio to do the aperture correction. This is insignificant compared to the scatter of the \LNII/\LIR-FIR color relation. However, the underestimation will reach 30\% for sources with $\log (f_{70}/f_{160})_{\rm out} > -0.2$. Therefore, it still worth using a color-dependent \LNII/\LIR\ relation to do the aperture correction since the underestimation is systematic.

\section{Results and Discussion}
\subsection{The \NIIab\ Emission as a SFR Indicator}                                                                    
\subsubsection{$L_{[{\rm N\,{\scriptsize II}}]}-$SFR Correlation}
\begin{figure*}[tbp]
\centering
\includegraphics[width=0.95\textwidth,bb = 5 120 508 362]{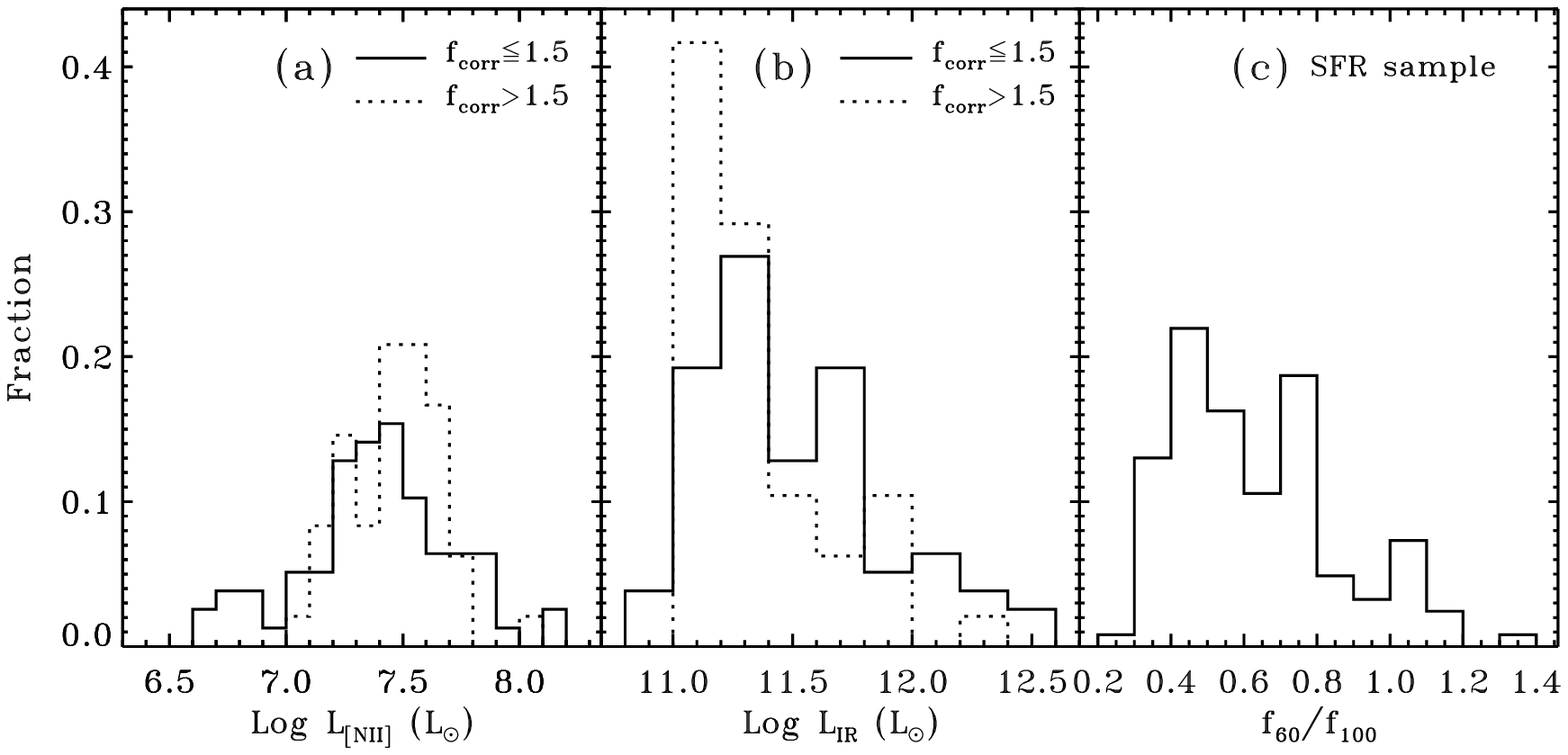}
\caption{Distributions of (a) and (b): \LNII\ and \LIR\ for the GOALS-FTS sample, respectively; (c): FIR color for the SFR sample. The solid and dotted histograms in panels (a) and (b) show the results for the $f_{\rm corr} \leq 1.5$ and $f_{\rm corr} > 1.5$ subsamples respectively.}
\label{Fighistnii}
\end{figure*}

To estimate the SFR of our sources we used the algorithm of Dale et al. (2007), e.g. ${\rm SFR}\,(M_\odot\,{\rm yr}^{-1})=4.5 \times 10^{-37}L_{\rm IR}\,({\rm W}) + 7.1 \times 10^{-37}\nu L_\nu\,(1500{\rm\,\AA})\,({\rm W})$, which takes into account dust obscuration by combining the {\it IRAS} IR and {\it GALEX} FUV fluxes. Here \LIR\ was calculated with the \iras\ four-band data. Without taking into account the dependence of SFR on the assumed initial mass function, the uncertainty in SFR from this composite calibrator is dominated by the uncertainty in the coefficient of the first term on the right side of the equation. Here we adopted an uncertainty of 40\% (e.g. Kennicutt \& Evans 2012), and it was propagated to the final SFR after taking a quadratic sum of all errors. For the GOALS-FTS sample, the FUV data were adopted from Howell et al. (2010), while for the normal galaxy and {\it ISO} samples, the FUV data were compiled from the literature (mainly from, e.g. Dale et al. 2007; Gil de Paz et al. 2007) and the {\it GALEX} data release GR7\footnote{http://galex.stsci.edu/GR6/\#5}. Since our SFR estimate relies on the availability of UV observations, we restricted our sample to 121 galaxies with available UV photometric data (hereafter SFR sample).

Before further analysis, however, it is instructive to check whether our dataset is capable of exploring a relation between \LNII\ and SFR. This is due to the fact that (1) for the GOALS-FTS sample, the aperture correction is essential the conversion of \LIR\ into \LNII; (2) The SFRs for the GOALS-FTS galaxies are dominated by \LIR. Therefore, \LNII\ and SFR may artificially correlate with each other even if they do not have an intrinsic relationship. As shown in Figures \ref{Fighistnii}a and \ref{Fighistnii}b, sources with $f_{\rm corr} \leq 1.5$ and $f_{\rm corr} > 1.5$ for our GOALS-FTS sample reside in similar phase space of \LNII\ and \LIR. In addition, very extended sources with $f_{\rm corr}>2.0$ only occupy a small fraction ($\sim$17\%) of the SFR sample. Therefore, we conclude that our dataset will not artificially make a correlation between \LNII\ and SFR.

\begin{figure*}[tbp]
\centering
\includegraphics[width=0.78\textwidth,bb = 32 10 490 540]{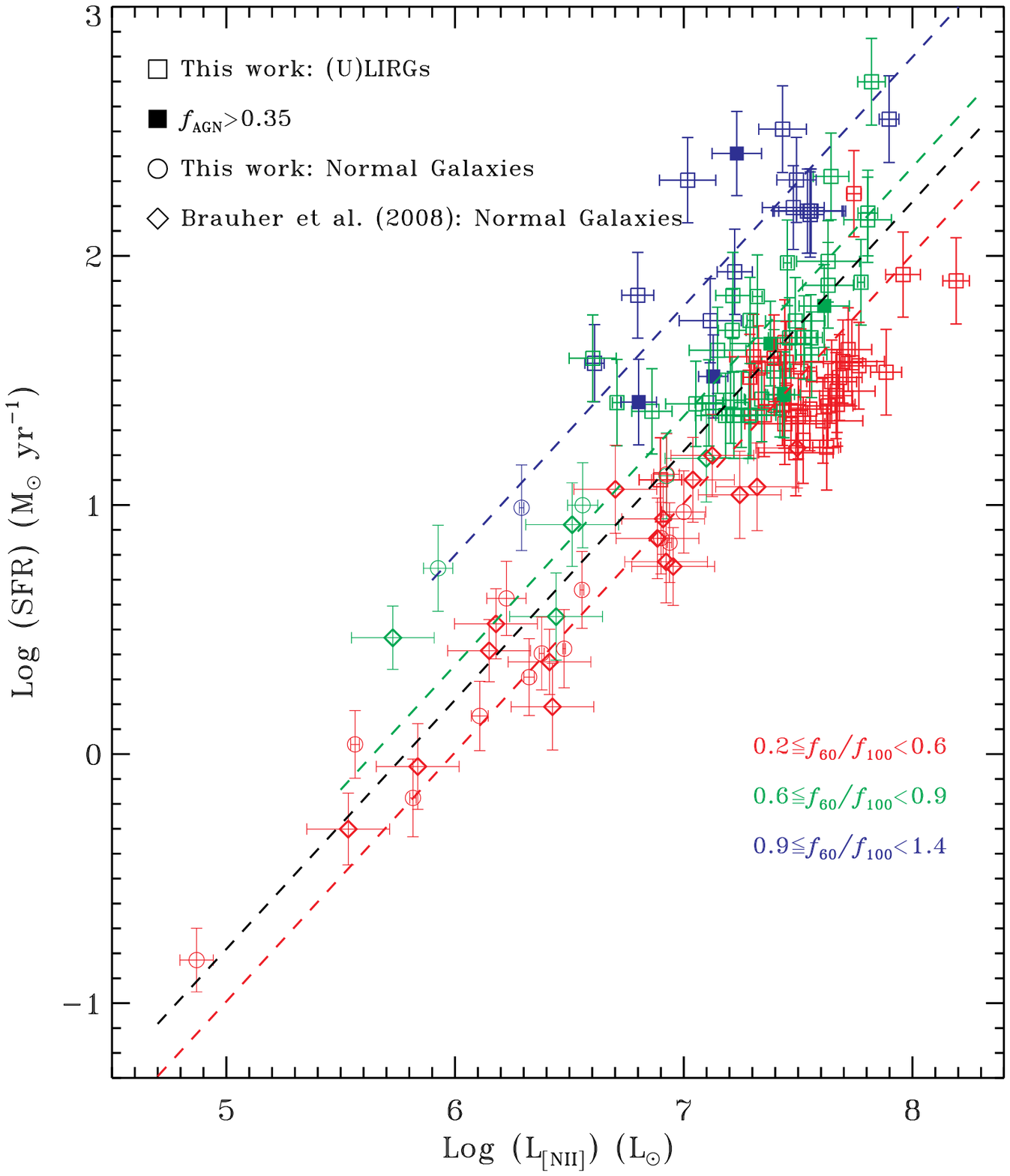}
\caption{Correlation between the \NIIab\ luminosity and SFR. The squares and circles are galaxies having \herschel\ observations; while the diamonds are galaxies from Brauher et al. (2008), whose \LNII\ were derived with the \NIIbc\ emission. The solid symbol indicates that the AGN contributes more than 35\% to the total bolometric luminosity, and are excluded from the fit. The points are color-coded according to their \fircolor. For each FIR color bin, the best-fit relation (slope of 1) is shown by the dashed-line, with the black line showing the relation for the entire sample.}
\label{Figsfr}
\end{figure*}

In Figure \ref{Figsfr}, we plot SFR against \LNII\ for both (U)LIRGs and normal galaxies. Squares represent the (U)LIRGs in the GOALS-FTS sample. Circles show normal galaxies observed by \herschel, whereas diamonds are the ISO sources from Brauher et al. (2008). The solid symbols in Figure \ref{Figsfr} indicate that the fractional contribution from a possible active galactic nucleus (AGN) to the bolometric luminosity, $f_{\rm AGN}$, is greater than 0.35. Here $f_{\rm AGN}$ was derived from a set of the mid-IR diagnostics based on [Ne\,{\sc v}]/[Ne\,{\sc ii}], [O\,{\sc iv}]/[Ne\,{\sc ii}], continuum slope, polycyclic aromatic hydrocarbon equivalent width, and the diagram of Laurent et al. (2000), following the prescriptions in Armus et al. (2007; see also Veilleux et al. 2009; Petric et al. 2011; Stierwalt et al. 2013). These galaxies are excluded from our fitting procedures for the \LNII-SFR relation(s). 

From Figure \ref{Figsfr} we can see that the scatter in \LNII$-$SFR relation becomes larger with the increase of SFR, consistent with Zhao et al. (2013). In this figure we demonstrate that the increase in scatter is traced to the individual galaxy colors (\fircolor). To isolate the color-dependence (and thus reduce the scatter) of the \LNII$-$SFR relation, we divide our sample galaxies into three sub-samples according to their \fircolor, i.e. a ``cold" one with $0.2\leq~ f_{60}/f_{100}< ~0.6$ (i.e. a blackbody temperature of $30 \lesssim T \lesssim 50$ K); a ``warm" one with $0.6\leq f_{60}/f_{100} < 0.9$ ($50 \lesssim T \lesssim 60$ K), and a ``hot" one with $0.9\leq f_{60}/f_{100} < 1.4$ ($60 \lesssim T \lesssim 90$ K). These color bins were chosen according to the FIR color distribution (three peaks in Figure \ref{Fighistnii}c) of our sample galaxies. Additional considerations for the separation of cold and warm/hot samples are that (1) Starburst galaxies usually have $f_{60}/f_{100}>0.55$ (Buat \& Burgarella 1998); (2) The turnover of the \LNII/\LIR-\fircolor\ relation happens at $f_{60}/f_{100}\sim 0.5$. 

\begin{figure*}[tbp]
\centering
\includegraphics[width=0.75\textwidth,bb = 12 40 502 716]{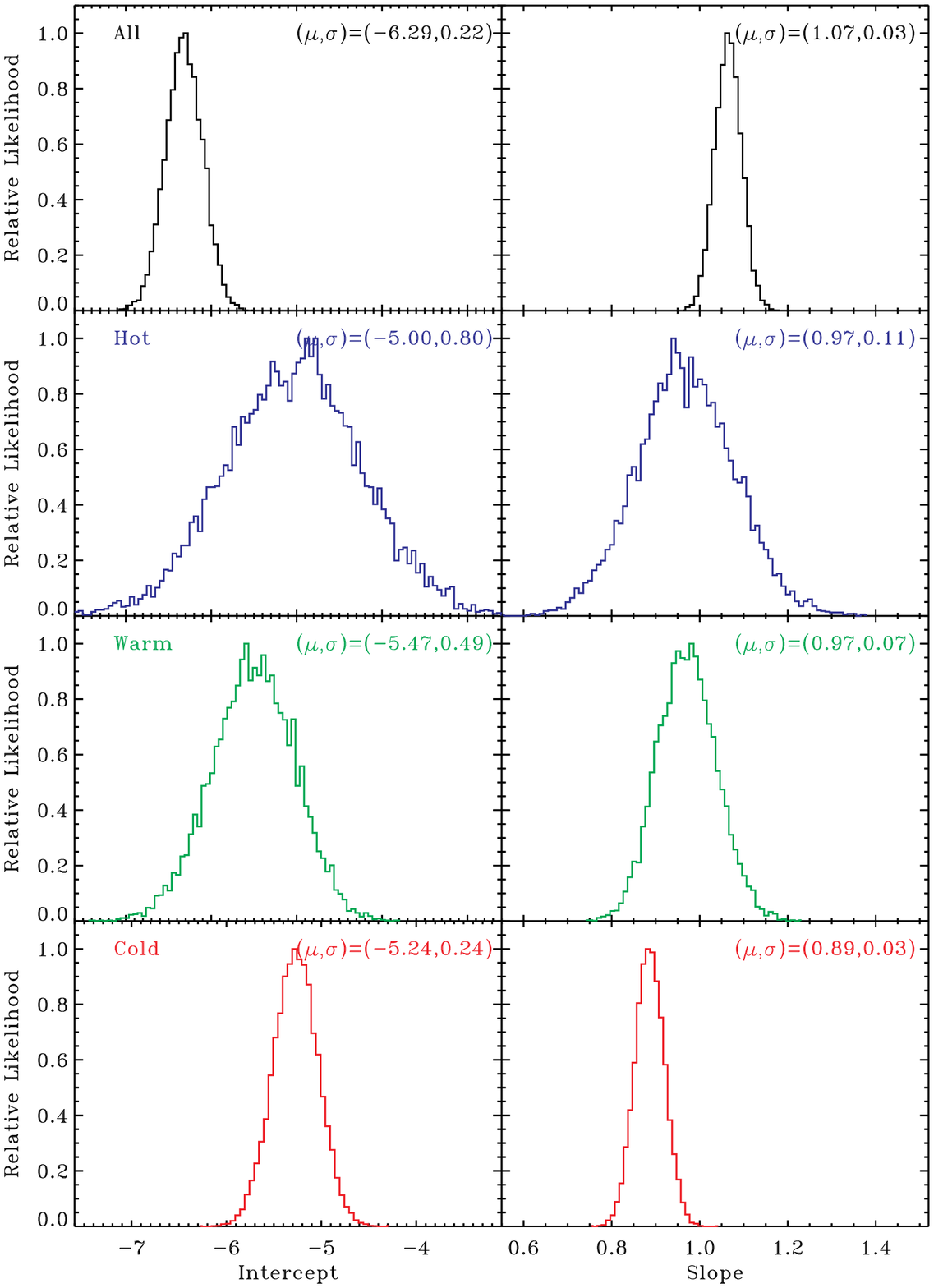}
\caption{Distributions of the fitted intercepts and slopes, which were obtained through the Monte Carlo simulations (see text for details), for each (sub-)samples. In each panel, the numbers in parentheses give the best-fit parameters for a Gaussian function.}
\label{Figmonte}
\end{figure*}

To investigate the relationship between \LNII\ and SFR, we fitted each sub-sample using a least-squares, geometrical mean functional relationship (Isobe et al. 1990) with a linear form, i.e.
\begin{equation}\label{eqniisfr}
\log\,{\rm SFR}\,(M_\odot\,{\rm yr}^{-1}) = a + b \log\,L_{\rm [N\,{\scriptsize \textsc{ii}}]}\,(L_\odot).
\end{equation}
to all galaxies except those having $f_{\rm AGN} > 0.35$. Using the same method, we also fitted the whole galaxy sample. To take into account the uncertainties both in \LNII\ and SFR, we used two independent approaches, which allow us to evaluate their reliabilities, to estimate the final coefficients ($a$, $b$) and associated errors in equation \ref{eqniisfr}. The first method (M1) was carried out by using a Monte Carlo simulation described as follows. Firstly, we generated a simulated sample by assuming a Gaussian error using the measured data points and uncertainties. Secondly, we fitted this sample using equation \ref{eqniisfr} and the geometrical mean method. Thirdly, we repeated the previous two steps 10000 times. The distributions of the fitted results from this process are shown in Figure \ref{Figmonte}. The second method (M2) is that the observed data points were fitted by using a weighted least-squares, geometrical mean regression. The weighting is defined after Williams et al. (2010), namely, $1/\sigma^2 \equiv 1/(b^2\sigma^2_{L_{\rm [NII]}}+\sigma^2_{\rm SFR})$, where $\sigma^2_{L_{\rm [NII]}}$ and $\sigma^2_{\rm SFR}$ are the errors in \LNII\ and SFR, respectively.

\begin{deluxetable*}{lccccccccc}
%updated using latest results;20150625
\centering
\tablecaption{Summary of the fitted Coefficients for Eq. (2)\label{fittedsum}}
\tablewidth{0pt}
\tabletypesize{\scriptsize}
\tablehead{
\colhead{\multirow{2}{*}{Sample}}&\colhead{\multirow{2}{*}{N}}&\colhead{Intercept}&\colhead{Slope}&\colhead{Scatter}&\colhead{\multirow{2}{*}{$\chi^2_{\rm red}$}}&\colhead{\multirow{2}{*}{Method}}&\colhead{\multirow{2}{*}{$\rho$}}&\colhead{\multirow{2}{*}{Sig}}&\colhead{Including}\\
&&\colhead{($a$)}&\colhead{($b$)}&\colhead{(dex)}&&&&&\colhead{ISO gal?}\\
(1)&(2)&(3)&(4)&(5)&(6)&(7)&(8)&(9)&(10)
}
\startdata
\multirow{4}{*}{Cold} & 62  & $-5.24$(0.24) & 0.89(0.03) & 0.18&--- &M1& 0.88 &$\gg5\sigma$&Yes\\
&{\bf 62}&$\mathbf{-5.17(0.22)}$&{\bf 0.88(0.03)}&{\bf 0.18}&{\bf 1.13}&{\bf M2}& {\bf 0.88} &$\mathbf{\gg5\sigma}$&{\bf Yes}\\
&46&$-5.26(0.38)$&0.90(0.05)&0.19&---&M1&0.59&$4.76\sigma$&No\\
&46&$-5.12(0.24)$&0.88(0.04)&0.18&1.29&M2&0.59&$4.76\sigma$&No\\
\hline
\multirow{4}{*}{Warm} & 39  & $-5.47$(0.49) & 0.97(0.07) & 0.25 &---&M1& 0.83 &$\gg5\sigma$&Yes\\
&{\bf 39}&$\mathbf{-5.53(0.53)}$&{\bf 0.98(0.07)}&{\bf 0.25}&{\bf 1.76}&{\bf M2}&{\bf 0.83} &$\mathbf{\gg5\sigma}$&{\bf Yes}\\
&35&$-5.65(0.64)$&1.00(0.09)&0.25&---&M1&0.78 &$\gg5\sigma$&No\\
&35&$-5.68(0.64)$&1.00(0.09)&0.25&1.83&M2&0.78 &$\gg5\sigma$&No\\
\hline
\multirow{2}{*}{Hot}  & 14  & $-5.00$(0.80) & 0.97(0.11) & 0.20 &---&M1& 0.61 &$2.51\sigma$&---\\
&{\bf 14}&$\mathbf{-5.04(0.85)}$&{\bf 0.98(0.12)}&{\bf 0.20}&{\bf 1.23}&{\bf M2}&{\bf 0.61} &$\mathbf {2.51\sigma}$&---\\
\hline
\multirow{4}{*}{All}  & 115 & $-6.29$(0.22) & 1.07(0.03) & 0.37 &---&M1& 0.72 &$\gg5\sigma$&Yes\\
&{\bf 115}&$\mathbf{-6.61(0.23)}$&{\bf 1.12(0.03)}&{\bf 0.37}&{\bf 3.23}&{\bf M2}&{\bf 0.72} &$\mathbf{\gg5\sigma}$&{\bf Yes}\\
& 95 & $-6.29$(0.33) & 1.07(0.03) & 0.38 &---&M1& 0.50 &$5.49\sigma$&No\\
& 95 & $-6.56$(0.26) & 1.11(0.03) & 0.38&3.63 &M2& 0.50 &$5.49\sigma$&No\\
\hline\noalign{\smallskip}
\multicolumn{10}{c}{Fixed slope}\\
\hline\noalign{\smallskip}
\multirow{2}{*}{Cold} & 62  & $-5.99$  & 1.0  & 0.22 &1.31 &M2& 0.88 &$\gg5\sigma$&Yes\\
& 46  & $-5.99$  & 1.0  & 0.22 &1.52 &M2& 0.59 &$4.76\sigma$&No\\
\multirow{2}{*}{Warm} & 39  & $-5.64$  & 1.0  & 0.25 &1.72 &M2& 0.83 &$\gg5\sigma$&Yes\\
& 35  & $-5.64$  & 1.0  & 0.25 &1.77 &M2& 0.78 &$\gg5\sigma$&No\\

Hot  & 14  & $-5.20$  & 1.0  & 0.20  &1.13 &M2& 0.61 &$2.5\sigma$&---\\
\multirow{2}{*}{All}  & 115 & $-5.78$  & 1.0  & 0.35 &3.33 &M2& 0.72 &$\gg5\sigma$&Yes\\
& 95 & $-5.77$  & 1.0  & 0.36 &3.71 &M2& 0.50 &$5.49\sigma$&No
%\vspace{-4pt}
\enddata
\tablecomments{Col. (1): Sample defined according to their \fircolor, i.e. $0.2\leq~ f_{60}/f_{100}< ~0.6$, $0.6\leq f_{60}/f_{100} < 0.9$, and $0.9\leq f_{60}/f_{100} < 1.4$ for ``Cold", ``Warm" and ``Hot", respectively. ``All" represents that the sample includes all galaxies. Col. (2): Number of sources in each sample. Cols. (3) and (4): Coefficients (and the $1\sigma$ uncertainties) between $L_{\rm [N\,{\scriptsize {\tiny \textsc{ii}}]}}$ and SFR, as defined in Eq. (2). Col. (5): rms scatter of the relation. Col. (6): Reduced chi-square. Col. (7): The method used to fit the relation. Cols. (8) and (9): The Spearman's rank correlation coefficient ($\rho$) and the level of significance (Sig; computed from the $p$-value using a Student's {\it t} distribution, which approaches the normal distribution as the sample size increases), respectively. Col. (10): Whether the \iso\ galaxies were included in the sample. None of the \iso\ galaxies has $0.9\leq f_{60}/f_{100} < 1.4$. The bold rows show the results that we used to calculate the $3\sigma$ range of the slope and discussed in more details in the main text.}
\end{deluxetable*}

Table \ref{fittedsum} lists the number of objects in each sample, the fitting coefficients, $1\sigma$ errors and scatters from both methods, for the \LNII$-$SFR relation. From the table we can see that M1 and M2 give consistent results. Hence, we only discuss the results from M2 hereafter. In Table \ref{fittedsum} we also show the Spearman's rank correlation coefficient ($\rho$, assessing how well an arbitrary monotonic function could describe the relationship between two variables), and the level of significance (Sig), which was computed from the $p$-value using a Student's {\it t} distribution. These $\rho$ and Sig indicate that there exists a very strong correlation between \LNII\ and SFR. 

On average, SFR scales with $L^b_{\rm [N\,{\scriptsize \textsc{ii}}]}$ with $b$ between 0.62 and 1.34 at $3\sigma$ significance. The slopes in the current work are consistent with the result ($0.95\pm0.05$) found in Zhao et al. (2013) within $1-2\sigma$ uncertainty ranges. The nearly linear relation between \LNII\ and SFR indicates that the power source of the \NIIab\ emission may be related to the details of the star formation processes that take place in each galaxy. Given such a strong correlation, and to reduce any systemic uncertainties caused by the sample itself (such as sample size, dynamic range, etc), we also fitted the \LNII$-$SFR relation with a fixed slope of 1. These results are also given in Table \ref{fittedsum}, and plotted in Figure \ref{Figsfr} as a dashed line for each sample. The reduced $\chi^2$ from both fixed and varying slopes, as listed in \ref{fittedsum}, agree with each other within $<15\%$, and thus the fitted results with a fixed slope of 1 are recommended to be used for computing SFRs.

Are the fitted relations sensitive to $R_{122/205}$ for our (sub-)samples? To further check this, we fitted the \LNII$-$SFR relations for the (sub-)samples by excluding the \iso\ galaxies and using the method M2. We found that the resultant slopes and intercepts only have tiny changes, as shown in Table \ref{fittedsum}. Therefore, we conclude that our results are not affected substantially by including the \iso\ galaxies.

\subsubsection{The Scatter in the \LNII$-$SFR Relation}
\begin{figure}[tbph]
\centering
\includegraphics[width=0.48\textwidth,bb = 58 4 479 712]{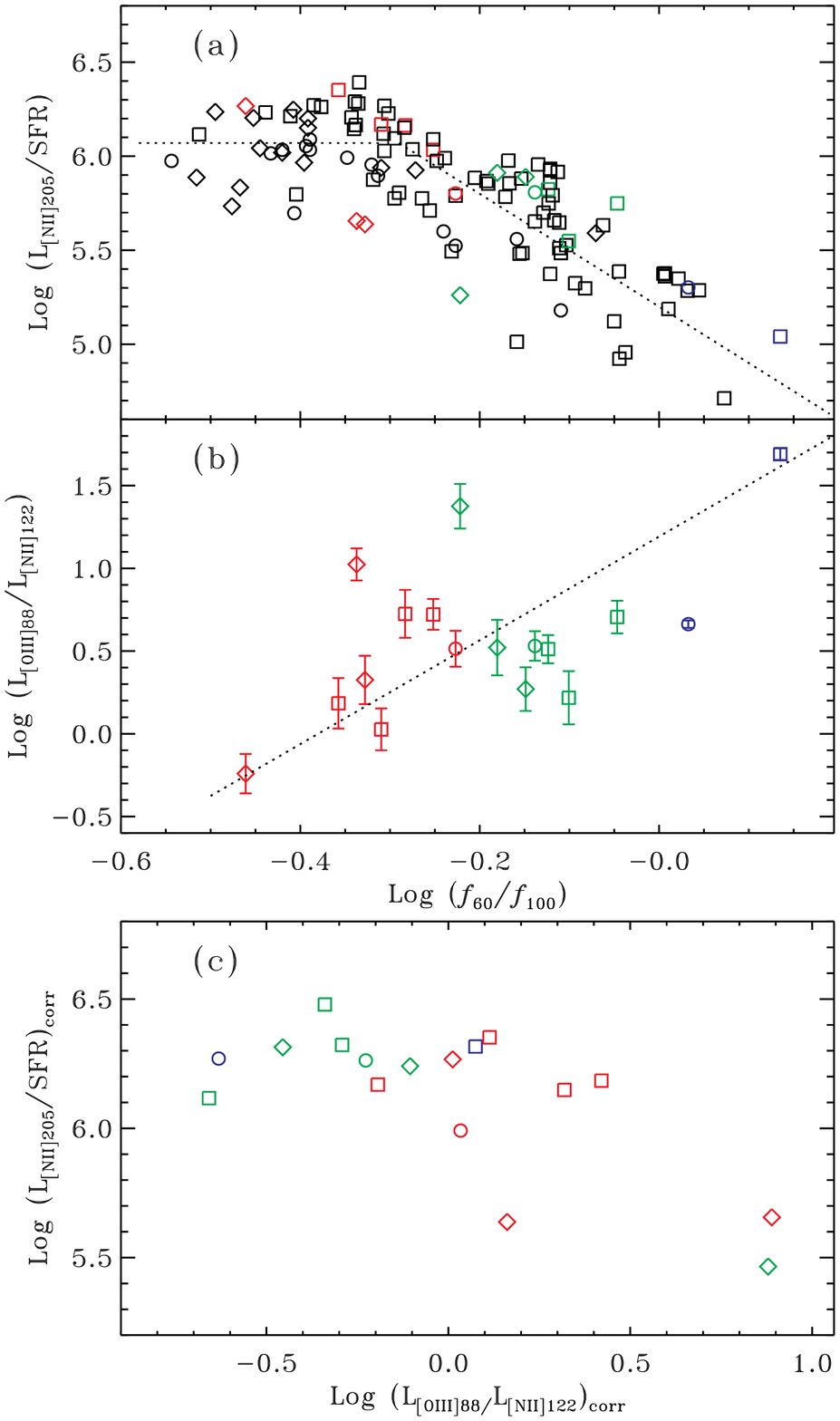}
\caption{(a) \LNIIab/SFR (in units of $L_\odot/(M_\odot\,{\rm yr}^{-1})$) plotted against \fircolor; (b) \OIIIFIR/\NIIbc\ vs \fircolor; and (c), \LNIIab/SFR vs \OIIIFIR/\NIIbc. The symbols are the same as in Figure \ref{Figsfr}. The points in panels (b) and (c), which have both \OIIIFIR\ and \NIIbc\ data, are outlined in panel (a) with colored symbols. In panel (a), the flat line is the mean value of $\log\,(L_{\rm [N\,{\scriptsize \textsc{ii}}]205\,\mu{\rm m}}/{\rm SFR})$ for the sources with $\log\,f_{60}/f_{100} \leq -0.29$, and the other line gives the result of a least-squares, geometrical mean fit to the subsample with $\log\,f_{60}/f_{100} \geq -0.29$. The line in panel (b) is $\log\,(L_{\rm [O\,{\scriptsize \textsc{iii}}]88\,\mu{\rm m}}/L_{\rm [N\,{\scriptsize \textsc{ii}}]205\,\mu{\rm m}})=1.2+3.1\log\,f_{60}/f_{100}$. The values of the x- and y-axis in panel (c) are obtained using equations (5) and (4), respectively.}
\label{Fighard}
\end{figure}

Table \ref{fittedsum} shows that the scatter of each sub-sample is a factor of $\sim$1.5 smaller compared to that of the full sample. This suggests that the color-dependence of the \NIIab\ emission contributes significantly to the total scatter of the \LNII$-$SFR relation. This is further confirmed by the following two checks: (1) A principal component analysis indicates that the FIR color accounts for 41\% of the total variance of the entire sample; (2) We simply normalized the \LNII\ by the galaxy FIR color, i.e., $\log\,L_{\rm [N\,{\scriptsize \textsc{ii}}]205\,\mu{\rm m,\,norm}}=\log\,L_{\rm [N\,{\scriptsize \textsc{ii}}]205\,\mu{\rm m}}+\log\,(f_{60}/f_{100})$, and then fitted the $\log\,L_{\rm [N\,{\scriptsize \textsc{ii}}]205\,\mu{\rm m,\,norm}}-{\rm SFR}$ relations with method M2 (varying slope) and calculated the scatters, which are 0.15, 0.23, 0.20 and 0.26 dex for the ``Cold", ``Warm", ``Hot" and ``All" samples, respectively. For the former three samples, these values are almost the same as those for the \LNII$-$SFR relation, while for the ``All" sample, it is reduced by a factor 1.3. If we ``corrected" the \LNII\ using the dotted lines in Figure \ref{Fighard}a, i.e.
\begin{equation}
	y=\left\{ \begin{array}{ll}
 6.07 &\mbox{ if $x \leq -0.29$} \\
  5.20 -3.00x &\mbox{ otherwise}
       \end{array} \right.
\end{equation}
where $y=\log\,(L_{\rm [N\,{\scriptsize \textsc{ii}}]205\,\mu{\rm m}}/{\rm SFR})$, in units of $L_\odot/(M_\odot\,{\rm yr}^{-1})$, and $x=\log\,(f_{60}/f_{100})$, their scatters would become 0.15, 0.21,0.22 and 0.18 dex respectively, and are comparable to the measurement uncertainty of 0.19 dex (median value of the entire sample).

As discussed in detail in Zhao et al. (2013), the scatter and/or color dependence in the \LNII$-$SFR (or \LNII$-$\LIR) relation is mainly due to the variation of ionization conditions in different galaxies. This is because the FIR color is tightly correlated with the ionization parameter, $U$ (Abel et al. 2009; Fischer et al. 2014; Comier et al. 2015). Adopting the \OIIIFIR-to-\NIIab\ flux ratio as an indicator of the hardness of the radiation field, Zhao et al. (2013) also suggested that the hardness variation can largely account for the scatter. However, the \OIIIFIR/\NIIab\ ratio is sensitive to electron density (Rubin 1985) since the levels emit these two lines have their critical densities differing by a factor of $>$10. Therefore, here we used the \OIIIFIR/\NIIbc\ ratio, which is insensitive to density, as a hardness indicator (Ferkinhoff et al. 2011) to further check the hardness effect. 

However, we note that the \OIIIFIR/\NIIbc\ ratio is only a good hardness indicator for a fixed $U$. It is correlated with $U$ at a given hardness (Cormier et al. 2015). Therefore, for the sample having both \OIIIFIR\ and \NIIbc\ data (hereafter ``OIII sample") we correct $\log\,(L_{\rm [N\,{\scriptsize \textsc{ii}}]205\,\mu{\rm m}}/{\rm SFR})$ (then the scatter changed from 0.36 dex to 0.28 dex), and the \OIIIFIR/\NIIbc\ ratios using the following equations, respectively:
%\setlength{mathindent}{0pt}
%\begin{flalign}
\begin{equation}
	  	\log\,(L_{\rm [N\,{\scriptsize \textsc{ii}}]205\,\mu{\rm m}}/{\rm SFR})_{\rm corr}  = \left\{ \begin{array}{ll}
 y &\mbox{ if $x \leq -0.29$} \\
  y-(5.20 -3.00x)+6.07 &\mbox{ otherwise}
       \end{array} \right.
\end{equation}
%\end{flalign}
and
\begin{equation}
  	\log\,(L_{\rm [O\,{\scriptsize \textsc{iii}}]88\,\mu{\rm m}}/L_{\rm [N\,{\scriptsize \textsc{ii}}]122\,\mu{\rm m}})_{\rm corr}= y_1-(1.2+3.1x)
\end{equation}
where $y=\log\,(L_{\rm [N\,{\scriptsize \textsc{ii}}]205\,\mu{\rm m}}/{\rm SFR})$, $y_1=\log\,(L_{\rm [O\,{\scriptsize \textsc{iii}}]88\,\mu{\rm m}}/L_{\rm [N\,{\scriptsize \textsc{ii}}]122\,\mu{\rm m}})$, and $x=\log\,(f_{60}/f_{100})$. Equation (5) is the result of a least-square fit to the \OIIIFIR/\NIIbc$-$\fircolor\ relation (see Figure \ref{Fighard}b). In this way, we may eliminate/reduce the effect of $U$ on both parameters. As shown in \ref{Fighard}c, there exists a weak correlation between $\log\,(L_{\rm [N\,{\scriptsize \textsc{ii}}]205\,\mu{\rm m}}/{\rm SFR})$ and $\log\,(L_{\rm [O\,{\scriptsize \textsc{iii}}]88\,\mu{\rm m}}/L_{\rm [N\,{\scriptsize \textsc{ii}}]122\,\mu{\rm m}})$, with the Spearman's rank coefficient $\rho=-0.5$ at a $2.1\sigma$ level of significance. The scatter of the OIII sample is reduced from 0.28 dex to 0.20 dex after further correcting for this hardness-dependence. Therefore, the variations of ionization parameter and hardness can be mainly responsible for the scatter in the \NIIab-SFR relation. Nevertheless, the OIII sample is small and biased towards higher $\log\,(L_{\rm [N\,{\scriptsize \textsc{ii}}]205\,\mu{\rm m}}/{\rm SFR})$ at a given FIR color (see Figure \ref{Fighard}a). A larger and more unbiased sample is needed in future studies.

\subsubsection{Comparisons with Other FIR Fine-structure Line-based SFR Indicators}
Other FIR atomic and ionic  fine-structure lines have been studied for their application as a SFR tracer (e.g. Farrah et al. 2013; De Looze et al. 2014). In comparison, our \NIIab\ line-based SFR tracer fares relatively well in terms of the overall uncertainty and the systematic dependence on the FIR color. Farrah et al. (2013) found that among the six lines ([O\,{\sc iii}]\,52\,$\mu$m, [N\,{\sc iii}]\,57\,$\mu$m, [O\,{\sc i}]\,63\,$\mu$m, \NIIbc, [O\,{\sc i}]\,145\,$\mu$m and [C\,{\sc ii}]\,158\,$\mu$m) studied, the [O\,{\sc i}] and \NIIbc\ are the most reliable tracers of SFR for a sample of ULIRGs, and the derived star formation rates for a given object are consistent to within a factor of three (0.48 dex). De Looze et al. (2014) found that the scatters are 0.46, 0.46 and 0.66 dex, for the \CII, \OIFIR\ and \OIIIFIR\ lines respectively. Whereas the \NIIab\ indicator has an accuracy of $<$0.4 dex, which can be further improved to $\sim$0.2 dex for a source with a known \fircolor. Therefore, the \NIIab\ line seems to have the smallest uncertainty although Farrah et al. (2013) did not report the exact  values of the scatter  in their SFR indicators. 

Many studies (i.e., Malhotra et al. 2001; Brauher et al. 2008; Graci\'{a}-Carpio et al. 2011; De Looze et al. 2014; Cormier et al. 2015) have demonstrated that most of the FIR fine-structure lines show some dependence on the dust temperature or FIR color. From ISM modeling the [O\,{\sc i}]\,63 and 145\,$\mu$m lines may be the most FIR color-insensitive lines at a {\it given} density, and the line-to-IR continuum luminosity ratios only vary about 0.3-0.5 dex for $0.3\leq f_{60}/f_{100}\lesssim2.2$ (Abel et al. 2009; Cormier et al. 2015). However, De Looze et al. (2014) showed that the ratio between \OIFIR-based and FUV+IR-based SFRs has a span of $\sim$1.3 dex for $f_{100}/f_{160}\sim0.5-2.1$ (equal to $f_{60}/f_{100}\sim0.1-1$ for a graybody SED assuming a dust emissivity index $\beta=1.5$), and the span is even larger for \CII-based SFRs. For the \NIIab\ line, the \LNII/SFR ratios vary about 1.7 dex (see Figure \ref{Fighard}a) for $0.2\leq f_{60}/f_{100}\leq 1.4$. Therefore, it seems that the \NIIab\ emission is not more sensitive to the FIR color than other lines, and we can conclude that the \NIIab\ emission is one of the most reliable SFR tracers among these fine-structure lines, in terms of its relatively small uncertainty.

\subsubsection{Applicability of the \NIIab\ Emission as a SFR Indicator at High-redshift}

The quantitative correlation between the \NIIab\ line and SFR for the local LIRGs implies that the \NIIab\ can be particularly a useful SFR tracer at high redshifts, because of the following reasons. Firstly, the \NIIab\ emission has an important advantage over the usual SFR calibrator, \LIR, for high-redshift galaxies: It might be relatively easier to be detected, since (a) it is resource intensive to construct a FIR SED to measure \LIR\ at high redshift (especially when $\lambda_{\rm rest} \leq 100\,\mu{\rm m}$, it needs very good weather to observe a source at $z=6$ even with ALMA, and gets harder for a lower redshift),  but \LIR\ derived with a single-band flux is very uncertain (see the review of Carilli \& Walter (2013)), and (b) for sources having very little dust, the IR continuum is very weak, but the FIR fine-structure lines might still be strong (e.g. Decarli et al. 2014; Capak et al. 2015; Ferkinhoff et al. 2015). 

Secondly, as already mentioned in \S1, the \NIIab\ line has an important advantage  over other shorter-wavelength FIR lines: it starts to fall into atmospheric sub/millimeter windows that have higher transmission at lower-$z$. Furthermore, as shown in Lu et al. (2015), the combination of the \CII\ 158 \mum\ [or CO(7-6)] and \NIIab\ emissions can be used to estimate the \fircolor\ (see their equations 4, 5 and 6), and hence we are able to obtain a more accurate SFR for those high-redshift galaxies if both lines are measured. 

However, for the application to high-$z$ galaxies, an issue that should be explored further is whether our SFR indicator depends on metallicity as some sources at higher redshifts might have significantly lower metallicities. Indeed, low-metallicity systems ($\sim$$0.25Z\odot$) at high-$z$ have already been observed (Capak et al. 2015). In our sample, there is one galaxy, ESO 350-IG 038 (Haro 11), which has a relatively low gaseous metallicity value of $\sim$0.48 $Z_\odot$ (Guseva et al. 2012){\footnotemark}{\footnotetext{In the current work, we adopt $12+({\rm O/H})_\odot=8.69$ (Asplund et al. 2009).}}, and $\log({\rm SFR}/L_{\rm [N\,{\scriptsize \textsc{ii}}]}) = -5.09\,(M_\odot\,{\rm yr}^{-1}\,L_\odot^{-1})$. We also found that another dwarf galaxy, He 2-10 ($Z=0.51 Z_\odot$; Guseva et al. 2011), has \herschel\ FTS observations (ObsID: 1342245083; PI: V. Lebouteiller). The derived $\log({\rm SFR}/L_{\rm [N\,{\scriptsize \textsc{ii}}]})$ for He 2-10 is $-5.28$ $(M_\odot\,{\rm yr}^{-1}\,L_\odot^{-1})$, where the SFR was calculated using the FIR and UV fluxes adopted from Leitherer et al. (2011). These two SFR/\LNII\ ratios are both within $1\sigma$ of the fitted relation at their colors $f_{60}/f_{100}>0.9$. Hence, it seems that the \NIIab\ emission is not very sensitive to metallicity. It might also be simply because the UV radiation field has a stronger influence on the \NII\ emission than metallicity does, given the fact that a lower metallicity will result in a stronger UV radiation field (e.g. Binney \& Merrifield 1998). 

In addition, Cormier et al. (2015) found that the average \LNIIbc/\LIR\ ratios for their dwarf galaxy sample with $7.0<12+{\rm (O/H)}<8.5$ and for the metal rich sample in Brauher et al. (2008) only differ by a factor of $\sim$2, which is very small compared to the effects on the \OIIIFIR\ line. Hence, the \NIIab\ line is not expected to strongly depend on metallicity down to $Z=1/15Z_\odot$ if we assume that \NIIab/\NIIbc\ is not a strong function of metallicity, which is reasonable. Therefore, the \NIIab\ emission is one of the FIR tracers least affected by metallicity effects, and can serve as a useful SFR indicator for high-redshift galaxies.
                    
\subsection{Local Luminosity Function and Star Formation Rate Density}
\subsubsection{Local LF of the \NIIab\ Emission}
\begin{figure*}[t]
\centering
\includegraphics[width=0.85\textwidth,bb = 1 38 500 308]{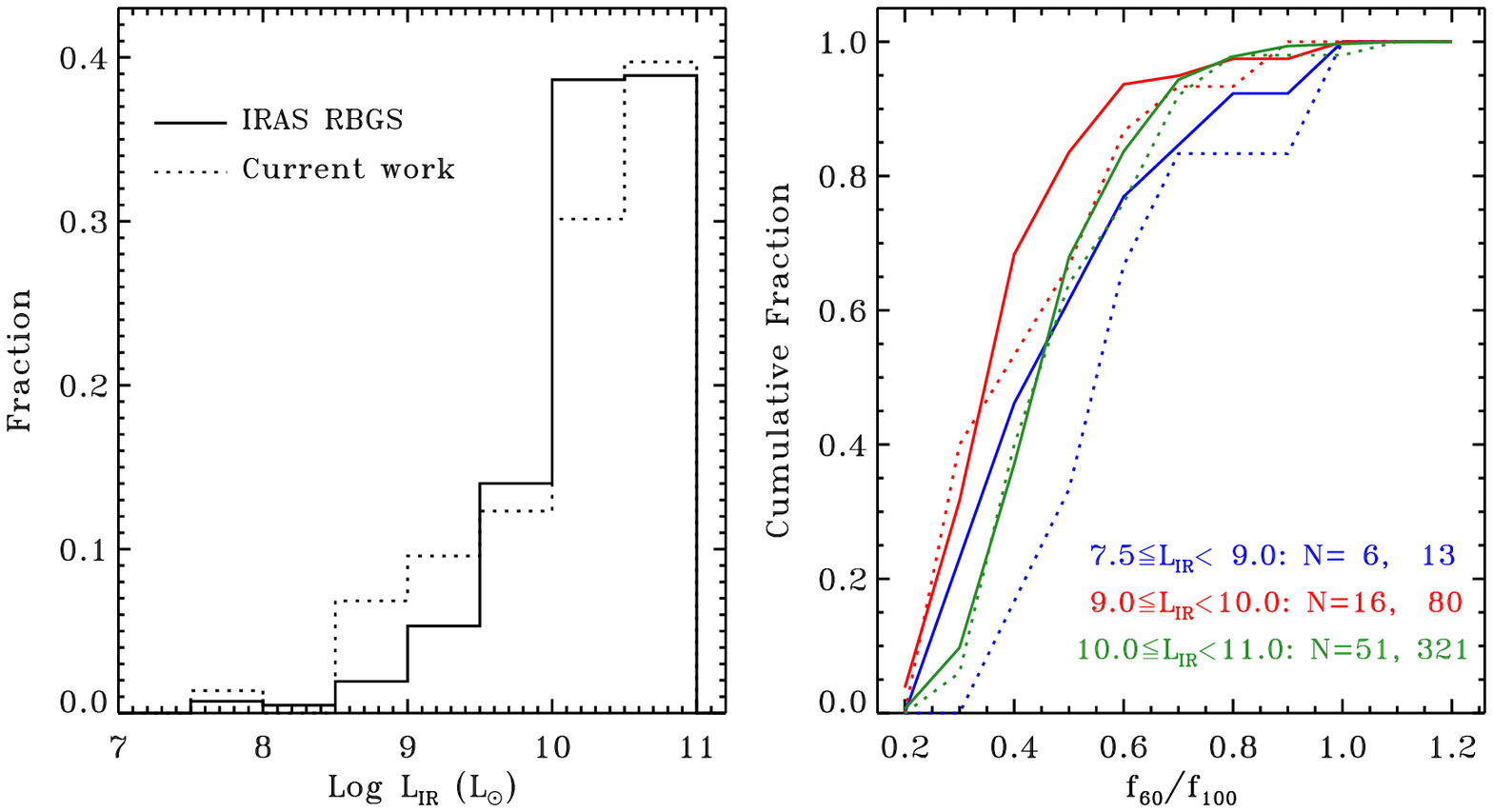}
\caption{Distributions of \LIR\ (left panel) and \fircolor\ (right panel) for the IRAS RBG sample (solid), which was used to construct the local infrared LF in Sanders et al. (2003), and our sample (dashed), which is used to derive the local LF of the \NIIab\ emission. In the right panel, the labels give the \LIR\ bins, and the number of sources in each bin (in the format of N=our sample, RBG sample). Only sources with $\log L_{\rm IR} < 11$ \lsolar\ are shown in the plots.}
\label{figirdist}
\end{figure*}

As shown in \S3.1, the \NII\ emission can be used as a SFR calibrator, which is particularly useful for high-redshift galaxies observable with submillimeter spectroscopic facilities such as ALMA. Therefore, besides its own evolution, studying the \NII\ LFs at different redshifts can probe the evolution of SFR density. In the following  we derive the local \NII\ LF using the data observed by \herschel\ and \iso. The final sample (hereafter LF sample) used to compute the local \NII\ LF contains 195 sources, in which 163 galaxies have a \NII\ detection, and the remaining 32 objects only have an upper limit on the \NII\ flux. For the \iso\ galaxies, \fNIIab\ was converted from \fNIIbc\ using the method described in \S2.2.2.

\begin{deluxetable}{ccc}
\centering
\tablecaption{\NII\ local luminosity function\label{niillf}}
\tablewidth{0pt}
\tabletypesize{\scriptsize}
\vspace{1mm}
\tablehead{
\colhead{$\log\,(L_{{\rm [N}\,{\tiny \textsc{ii}}]}/L_\odot)$}&$\log\,\left[\phi/({\rm Mpc^{-3}}\,{\rm dex}^{-1})\right]$&1$\sigma$ error%\vspace{1mm}
}
\startdata
5.0  & $-2.17$ $(-1.96)$ & $-2.50$ $(-2.46)$\\
5.5  & $-2.60$ $(-2.87)$ & $-2.85$ $(-3.16)$\\
6.0  & $-2.73$ $(-2.19)$ & $-3.10$ $(-2.91)$\\
6.5  & $-2.99$ $(-2.85)$ & $-3.47$ $(-3.20)$\\
7.0  & $-3.39$ $(-3.50)$ & $-3.88$ $(-3.78)$\\
7.5  & $-4.22$ $(-4.29)$ & $-4.76$ $(-4.59)$\\
8.0  & $-5.66$ $(-5.66)$ & $-6.11$ $(-6.10)$\\
\hline\hline\noalign{\smallskip}
&\multirow{1}{*}{Fitted Parameters{\tablenotemark{a}}}&\\
\hline\noalign{\smallskip}
%\vspace{-3pt}
\multirow{1}{*}{$\log\,\left[\phi^*/({\rm Mpc^{-3}}\,{\rm dex}^{-1})\right]$} & \multirow{1}{*}{$\log\,(L^*/L_\odot)$}&\multirow{1}{*}{$\alpha$}\vspace{1mm}\\
\hline\noalign{\smallskip}
%\vspace{-3pt}
\multirow{1}{*}{$-3.83\pm0.09$}& \multirow{1}{*}{$7.36\pm0.04$}  & \multirow{1}{*}{$-1.55\pm0.06$}
%\vspace{-4pt}
\enddata
\tablecomments{Numbers in parentheses give the results excluding \iso\ galaxies whose \LNII\ are converted from \LNIIbc.}
\tablenotetext{a}{For the Schechter (1976) function.}
\end{deluxetable}

Although our GOALS-FTS sample is IR flux-limited, it is not a \NII\ flux-limited sample. Also, we have added more normal galaxies into the study to improve the extent of our GOALS-FTS sample. Therefore, considering the breadth of our sample, we are able to use the so-called bivariate method (see, e.g., Xu et al. 1998) to construct the local \NII\ LF, in which we transferred the local IR LF of the \iras\ Revised Bright Galaxy Sample (RBGS) studied in Sanders et al. (2003) to a local \NII\ LF utilizing the \LNII/\LIR\ ratio versus the \LIR\ relation. However, as demonstrated in Figure \ref{Figcolordep} , the \LNII/\LIR\ ratio is dependent on the FIR color. Therefore, we further checked whether ours can be a representative sample of the \iras\ RBGS. To this end, we plot the fractional distribution of \LIR, and the cumulative fraction of \fircolor\ in three \LIR\ bins, in the left and right panels of Figure \ref{figirdist} , respectively. Due to the fact that our GOALS-FTS sample is complete and unbiased for galaxies of $L_{\rm IR} > 10^{11}$ \lsolar, we plot only the sources with $\log L_{\rm IR} < 11$ \lsolar\ in Figure \ref{figirdist}. 

In Figure \ref{figirdist} we can see that, in general, our sources have similar FIR color distributions to the \iras\ RBGS galaxies, albeit the discrepancy is relatively larger in the low luminosity bins, which could be attributed to the small number statistics involved. Using the $ad.test$ task in the {\it kSamples} package within {\bf R}, we performed an Anderson-Darling test (Adderson \& Darling 1952) to these two samples, and got a $p$-value of 0.17. Therefore, we do not reject the null hypothesis that these two samples arise from a common distribution. Although the discrepancy between the cumulative fractions of the \fircolor\ distribution for the least luminous \LIR\ bin with $f_{60}/f_{100} < \sim 0.6$ appears large, the \LNII/\LIR\ ratio is much less sensitive to the FIR color for these sources (see Figure \ref{Figcolordep}). Nevertheless, one should keep this caveat in mind when using the local \NII\ LF derived in the current work.

\begin{figure*}[t]
\centering
\includegraphics[width=0.75\textwidth,bb = 22 10 478 335]{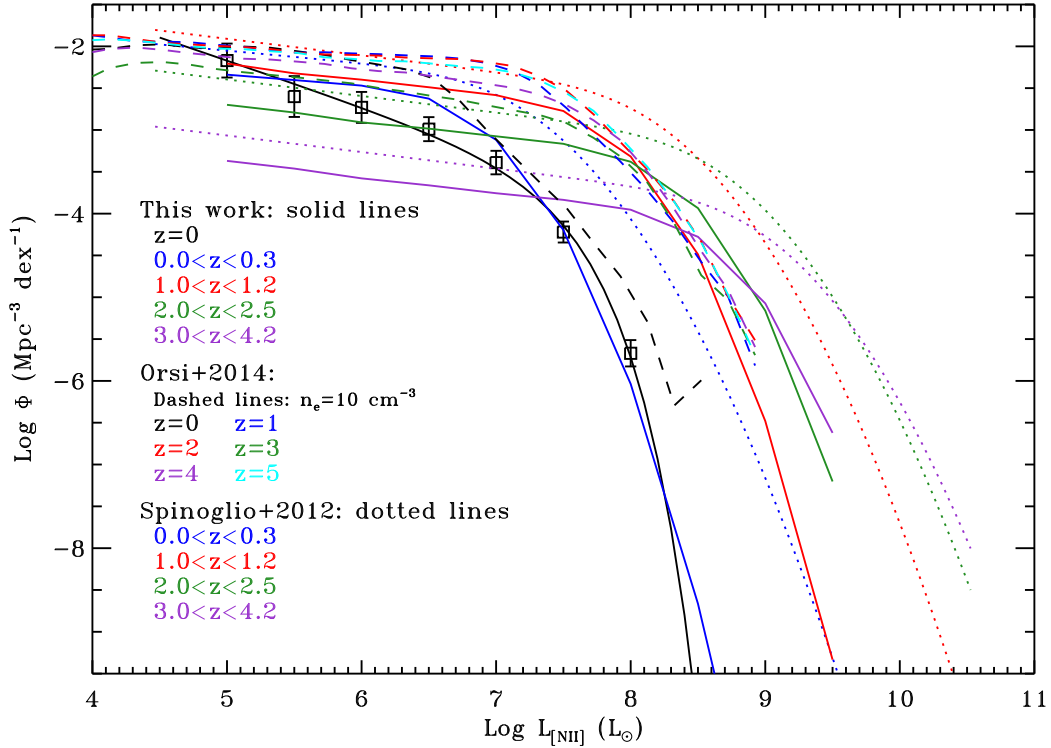}
\caption{The \NIIab\ local differential luminosity function (squares; Table \ref{niillf}). The black solid line shows the smooth fit with the Schechter (1976) function to the data. In addition, we plot the predicted LFs of \NIIab\ at different redshifts, derived with the method in section 3.2.3 (solid colorful lines) and the method presented in Spinoglio et al. (2012a; dotted lines) from the Gruppioni et al. (2013) IR LFs, as well as the modeling results from Orsi et al. (2014; dashed lines). See the text for details.}
\label{figlf}
\end{figure*}

The algorithm and the formulation used in this work are the same as those presented in \S3 of Xu et al. (1998)\footnote{The correct versions of equations (14) and (15) of Xu et al. (1998) should have read ${\rm Covar}(F_{i-1},F_i)=F_i {\rm Var}(F_{i-1})/F_{i-1}$, and ${\rm Var} \left[ \log (L_{15\,\mu{\rm m}}) = L_i\right] = \sum\limits_{j}\big\{ {\rm Var} (P_{i,j})\rho\left[ \log (L_{25\,\mu{\rm m}})=L_j\right] ^2 \allowbreak + P^2_{i,j} {\rm Var} \left[ \rho \log (L_{25\,\mu{\rm m}})=L_j\right] \big\} (\Delta_j/\delta_i)^2$, respectively.}, which were based on the Kaplan-Meier estimator (Kaplan \& Meier 1958) and therefore can take into account the information contained in upper limits. Here we adopted the local infrared bolometric (i.e. $L_{\rm IR}(8-1000\,\mu{\rm m})$) LF derived by Sanders et al. (2003). In Table \ref{niillf} we list the derived local \NII\ LF with bin width $\delta \log\,(L_{[{\rm N\,\textsc{ii}}]})=0.5$\,\lsolar, and the associated uncertainties. In Figure \ref{figlf} we plot this new local \NII\ LF, as well as the best fitted Schechter (1976) function, i.e.,
\begin{equation}
\phi(L)dL=\phi^*\left( \frac{L}{L^*}\right) ^\alpha \exp\left( -\frac{L}{L^*}\right)  d\left( \frac{L}{L^*}\right) 
\end{equation}
where $L$ is the galaxy luminosity, and $\phi(L)dL$ is the number density of galaxies in luminosity range $L+dL$. The parameters, $L^*$, $\phi^*$ and $\alpha$, determined from the fit, describe the characteristic luminosity, the normalization factor at $L^*$, and the slope of the LF at faint luminosities, respectively. The fitted parameters are given in Table \ref{niillf}.
%%%

As described earlier, we included the \iso\ sources into our analysis to make our sample larger and more representative in the low \LIR\ bins. To check how our results will change, we also derived the \NIIab\ LF by excluding the \iso\ galaxies (see Table \ref{niillf}), and found that the differences are pretty large ($\sim$$0.2-0.5$ dex) for $\log\,(L_{\rm [N\,{\scriptsize \textsc{ii}}]}/L_\odot)<6.5$. This is because only 13 galaxies with $L_{\rm IR}<10^{10.5}\,L_\odot$ have \NIIab\ data. Furthermore, \LIR\ of these sources do not cover the full range of the RBG sample. However, the inclusion of the \iso\ galaxies in our analysis is not unreasonable since the uncertainty of $\sim$0.2 dex in the resultant \LNII\ (converted from \LNIIbc) is much less than the luminosity bin we used to derived the LF. Nevertheless, we caution the reader to keep this caveat in mind when interpreting/using our results.

%%%
\subsubsection{Local Star Formation Rate Density}
It appears that the Schechter function provides a good fit to our data, with a reduced Chi-square of $\chi^2=0.31$ (see also Figure \ref{figlf}). We can further use this LF to compute the volume-averaged SFR density (SFRD) for the local Universe, by integrating the fitted Schechter function and multiplying by the SFR calibration provided in \S3.1, i.e.
\begin{equation}
\begin{split}
\log \dot{\rho}_{\rm SFR} (M_\odot\,{\rm yr}^{-1}\,{\rm Mpc^{-3}})=\log \left(\int_0^\infty\phi(L)LdL\right)-5.78\\
=\log\,\phi^* L^* \Gamma(2+\alpha)-5.78=-1.96, 
\end{split}
\end{equation}
where the adopted \LNII$-$SFR relation is based on the total sample with a fixed-slope (see the last line in Table \ref{fittedsum}). The quoted uncertainty in $\dot{\rho}_{\rm SFR}$ is about 0.11 dex when only the fitted errors of the local LF are taken into account. However, in the worst case scenario, i.e., the full amount of uncertainty in the SFR calibrator propagates to the final error, the resulting uncertainty in $\dot{\rho}_{\rm SFR}$ could be as large as 0.37 dex. 

The local SFRD obtained here is in good agreement with those of Gallego et al. (1995) and Yun et al. (2001), which give $\log \dot{\rho}_{\rm SFR}$ of $-2.02\pm0.2$ and $-1.86\pm0.14\,M_\odot\,{\rm yr}^{-1}\,{\rm Mpc^{-3}}$, respectively. These values were derived by assuming the same cosmological model as in this paper, and using the Kennicutt \& Evans (2012) SFR calibrators of H$\alpha$ and IR respectively. Our \NIIab-based $\dot{\rho}_{\rm SFR}$ is also consistent with other UV- and IR-based SFRD (e.g. Robotham \& Driver 2011; Takeuchi et al. 2003), as well as the best-fitting results given by Madau \& Dickinson (2014).

\subsubsection{Predictions and Comparisons of the \NIIab\ LFs at Different Redshifts}

 Our results provide a local benchmark for studying the evolution of the \NIIab\ emission and $\dot{\rho}_{\rm SFR}$. Here we further predict the \NIIab\ LFs at higher redshifts using the method described below. We will also compare our results with those from simulations given by Orsi et al. (2014), and those derived with the method presented in Spinoglio et al. (2012a). 

While nearby (U)LIRGs show a \NII\ deficit (relative to IR continuum emission) as compared to local normal galaxies (e.g. Graci\'{a}-Carpio et al. (2011), observations at high-$z$ indicate that not all of the high-redshift (U)LIRGs show this similar deficit (Ferkinhoff et al. 2011, 2015; Combes et al. 2012; Decarli et al. 2012, 2014; Nagao et al. 2012). This is expected if these (U)LIRGs at high redshifts have lower FIR colors than the local galaxies of comparable luminosities. Indeed, Casey et al. (2014 and references therein) shows that, for a fixed \LIR, the average dust temperature of a galaxy decreases as redshift increases. This implies that the ``turning" point in Figure \ref{Figcolordep} might occur at a higher characteristic \LIR\ if the trend shown in that figure remains independent of redshift. Inspired by these observations, here we have derived the \NIIab\ LFs at different redshifts from the IR LFs given by Gruppioni et al. (2013), by making the following three assumptions:
\begin{enumerate}
	\item The \LNII/\LIR\ ratio has the same $T_{\rm dust}$-dependence (cf. Eq. (1)) at any redshift;
	\item The \LIR$-\lambda_{\rm peak,\,dust}$ relation has the same power law (i.e., $\lambda_{\rm peak,\,dust}=K L_{\rm IR}^{-0.06}$, where $\lambda_{\rm peak,\,dust}$ is the SED peak wavelength and is inversely proportional to $T_{\rm dust}$) at any redshift, but with the scaling factor $K$ depending on redshift (Casey et al. 2014);
	\item In order to determine the scaling factor $K$ in (2), we used a fixed $T_{\rm dust}$ at $L_{\rm IR} = L^*$, where $L^*$ is the characteristic luminosity in the luminosity function defined by Gruppioni et al. (2013) and is an increasing function of redshift. Here we related $T_{\rm dust}$ to $\lambda_{\rm peak,\,dust}$ using a greybody with $\beta=1.5$ and $\tau=1$ at 200\,\mum\ (e.g. Casey 2012). Generally, the measured $T_{\rm dust}$ at $L^*$ for $z=0$ and higher redshifts (Casey et al. 2014) are consistent with our assumption here.
\end{enumerate}

The LFs predicted by our model calculation at various redshifts are shown in Figure \ref{figlf} with solid colored lines. Our measured \NIIab\ LF ($z=0$; solid black line) from Table \ref{niillf} generally agrees (within $1-3\sigma$) with the one at $0<z<0.3$ predicted by our method. The difference may be caused by different IR LFs (Sanders et al. LF vs Gruppioni et al. 2013 LF), and/or methods employed. For comparison, we also plotted the results (dotted lines) derived with the Spinoglio et al. (2012a) method, i.e., using an empirical \LIR$-$\LNIIbc\ relation calibrated locally (here we have converted the \LNIIbc\ to \LNIIab\ by assuming $R_{122/205}=1$ ($n_e\sim10$\,cm$^{-3}$); it would not change the LF shapes when using a different $R_{122/205}$). We can see that, both our measured and predicted LFs decline faster when $L_{\rm [N\,{\scriptsize \textsc{ii}}]205\,\mu m} > L^*$ than that based on the Spinoglio et al. (2012a) method. This difference is expected since our \LNII/\LIR\ ratios for warmer (and more luminous) galaxies are smaller than Spinoglio et al. (2012a).

In Figure \ref{figlf}, the dashed lines show the model predictions for $z=0-5$ from Orsi et al. (2014) with $n_e=10$\,cm$^{-3}$, which were obtained by using a semi-analytical model of galaxy formation and photoionization code. Except for one or two points, their model predicted LF at $z=0$ is consistent with our results within $1-3\sigma$. However, the Orsi et al. (2014) LF at $z=0$ is systematically larger than ours. The difference might be reduced by adopting a larger $n_e$ and/or varying $n_e$. From the figure we can see that, for higher redshifts, although the Orsi et al. (2014) results in the low-\LNII\ bins are much different from our predicted LFs, they show better agreement in the high-luminosity bins. The most probable reason for this disagreement is that too few points were used to construct the IR LFs at higher redshifts (Gruppioni et al. 2013), which could result in large uncertainties in the IR (\NIIab) LFs in the low-luminosity bins.

\subsection{Electron Density of the Ionized Gas in (U)LIRGs}
\begin{deluxetable*}{lcccc}
\centering
\tablecaption{Observed and derived parameters for galaxies observed at both \NII\ lines\label{tablene}}
\tablewidth{0pt}
\tabletypesize{\scriptsize}
\tablehead{
\colhead{Galaxy}&\colhead{\NIIbc\tablenotemark{a}}&\colhead{\NIIab}&\colhead{$R_{122/205}$}&\colhead{$n_e$\tablenotemark{b} (cm$^{-3}$)}
}
\startdata
IRAS17208-0014\dotfill& $<30$ & $5.4\pm0.8$&$<5.52$&$<334.2$  \\
IRASF10565+2448\dotfill& $<16$& $3.7\pm0.7$&$<4.38$& $<203.5$ \\
Mrk 231\dotfill&$  4.1\pm  0.5$&$2.9\pm0.5$&$1.40\pm0.29$&$26.7\pm1.1$\\
Mrk 273\dotfill&$  <14$& $3.7\pm0.4$&$<3.83$&$< 158.0$\\
Mrk 331\dotfill&$  23\pm6$&$10.0\pm0.3$&$2.30\pm0.61$&$65.8\pm3.1$\\
NGC 0695\dotfill&$15\pm3$&$15.2\pm2.7${\tablenotemark{c}}&$0.99\pm0.26$&$11.7\pm0.5$\\
NGC 1068\dotfill&$  363\pm30$&$304.6\pm124.0${\tablenotemark{c}}&$1.19\pm0.49$&$19.0\pm1.3$\\
NGC 1614\dotfill&$  29\pm6$&$15.6\pm3.5${\tablenotemark{c}}& $1.86\pm0.56$&$45.6\pm1.6$\\
NGC 2388\dotfill&$  36\pm7$&$20.5\pm3.5${\tablenotemark{c}}& $1.75\pm0.45$&$41.1\pm2.6$\\
NGC 3256\dotfill&$  123\pm25$&$90.9\pm23.3${\tablenotemark{c}}& $1.35\pm0.44$&$25.0\pm0.2$\\
NGC 4194\dotfill&$  <19$& $9.9\pm0.9$& $<1.92$ & $<48.4$\\
NGC 6240\dotfill&$23.0\pm2.7$&$15.3\pm1.5$&$1.51\pm0.23$&$31.0\pm1.5$\\
NGC 6286a\dotfill&$  17\pm5$&$33.6\pm5.3${\tablenotemark{c}} &$0.51\pm0.17$&$<1$\tablenotemark{d}\\
NGC 7552\dotfill&$  76\pm10$&$88.8\pm27.1${\tablenotemark{c}}&$0.86\pm0.28$&$7.1\pm0.1$\\
NGC 7771a\dotfill&$  47\pm10$&$41.8\pm9.7${\tablenotemark{c}}&$1.12\pm0.36$&$16.6\pm0.3$\\
UGC 02238\dotfill& $  13\pm4$ & $20.5\pm2.2${\tablenotemark{c}} & $0.56\pm0.21$&$<1$\tablenotemark{d}\\
\vspace{-4pt}
\enddata
\tablecomments{Fluxes are in units of $10^{-17}\,{\rm W\,m}^{-2}$.}
\tablenotetext{a}{All data, except Mrk 231, are adopted from Brauher et al. (2008). The data for Mrk 231 are adopted from Fischer et al. (2010).}
\tablenotetext{b}{The quoted error is obtained from the Monte Carlo simulation by assuming a Gaussian noise distribution.}
\tablenotetext{c}{Aperture corrections have been applied.}
\tablenotetext{d}{The observed ratios are smaller than (or approach) the theoretical lower limit.}
\end{deluxetable*}

\begin{figure}[t]
\centering
\includegraphics[width=0.48\textwidth,bb = 66 38 428 290]{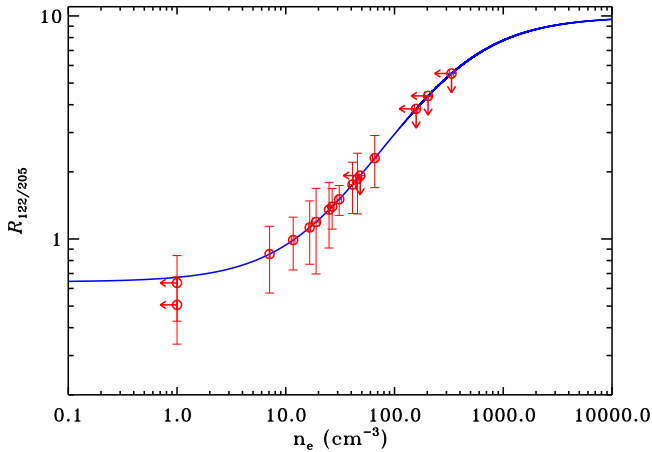}
\caption{Theoretical relation between the electron density ($n_e$) and \NIIbc\ to \NIIab\ ratio ($R_{122/205}$). Our measurements are shown by circles, whereas upper limits are indicated by arrows. Two data points with the smallest $R_{122/205}$ are shifted to $n_e=1$ cm$^{-3}$ since their $R_{122/205}$ are smaller than (or equal to) the theoretical limit.}
\label{figr21}
\end{figure}

In Figure \ref{figr21}, the solid curve is the theoretically expected $R_{122/205}$ as a function of electron density. This was derived by assuming collisional excitation by electrons from the ground-state levels of N$^+$, and using the collision strengths calculated with the fitted results (Draine 2011b) of Hudson \& Bell (2005) with $T=8000$ K. For galaxies in our GOALS-FTS sample, sixteen have \NII\ 122 $\mu$m data (12 detections and 4 upper limits), as listed in Table 2. Among these 16 galaxies, Mrk 231 has PACS observations and the flux has been measured by Fischer et al. (2010), whereas the other sources were observed by {\it ISO} and the fluxes are adopted from Brauher et al. (2008). To derive $n_e$, we match our observed $R_{122/205}$ (circles in Figure \ref{figr21}) to the theoretically expected curve. The derived $n_e$ ranges from $<1$ to tens cm$^{-3}$ (also see Table \ref{tablene}), and has a median value of 22 cm$^{-3}$ (corresponding $R_{122/205}=1.27$) for the 12 galaxies having \NII\ 122 \mum\ detections. This median $R_{122/205}$ for our sample is a bit larger than those of the Milky Way (0.9; Wright et al. 1991; Bennett et al. 1994) and the nearby spiral galaxy M51 ($0.8-0.9$; Parkin et al. 2013) as well as Cen A (0.8; Parkin et al. 2014), which might suggest that the density of the ionized medium traced by the \NII\ emission is higher in these (U)LIRGs. However, more data are needed to reach a solid conclusion since the current sample is too small and not representative of the (U)LIRG population.

Our results indicate that the \NIIab\ emission is dominated by the low-density ionized gas even in (U)LIRGs. This might be due to the fact that, for dusty \HII\ regions, the ionization parameter is so high in the high-density regime ($U$ is almost proportional to density; see, e.g. Draine 2011a), such as (Ultra-)compact \HII\ regions, that (1) the UV photons are preferentially absorbed by dust (Voit 1992; Bottorff et al. 1998; Draine 2011a), and (2) Most N$^+$ become N$^{++}$; producing much less \NII\ emissions (e.g. Abel et a. 2009; Fischer et al. 2014). Hence, most of the \NII\ emissions should come from the low-density surrounding halos of these dense \HII\ regions in (U)LIRGs. Indeed, high angular resolution ALMA observations of (U)LIRGs have revealed that compact regions of high gas density ($\sim 10^4\, {\rm cm}^{-3}$) are surrounded by more extended low density regions (Xu et al. 2014; 2015). This scenario is also consistent with our finding that sources with warmer \fircolor\ colors generally have lower \LNII-to-SFR ratios.

\section{Summary}
We present \herschel\ SPIRE/FTS observations of the \NIIab\ emission for an IR flux-limited sample of 122 (U)LIRGs. Combining them with SPIRE/FTS mapping observations of nearby normal galaxies, along with GALEX UV observations, PACS and \iras\ photometry, and \iso\ \NIIbc\ spectra, we have studied the dependence of the \LNIIab/\LIR\ ratio on the FIR color and \LNIIab-SFR correlation. We also derived the \NIIab\ luminosity function and SFR density for the local Universe, as well as investigated the electron densities for a sub-sample of (U)LIRGs. Our main results are summarized in the following.
\begin{enumerate}
\item
The \LNII/\LIR\ ratio depends on the FIR colors, while the dependence is much more sensitive when $\log\,(f_{70}/f_{160})> -0.2$ (equal to $f_{60}/f_{100} > 0.46$) than $\log\,(f_{70}/_{f160}) \lesssim -0.2$. 

\item
\LNII\ shows a strong, almost linear correlation with SFR, although the intercept of such correlation varies slightly according to the {\it IRAS} $f_{60}/f_{100}$. This agrees with the conclusion in Zhao et al. (2013) that \NIIab\ emission can be a useful SFR indicator. The overall uncertainty is about 0.4 dex, whereas the accuracy could be improved to $\sim$0.2 dex if we know the $f_{60}/f_{100}$ independently, and use the color-dependent \LNII-SFR relation to derive the SFR (see Table \ref{fittedsum}). 

\item
Using the bivariate method and combining the \iso\ \NIIbc\ data, which were converted to \NIIab\ emission based on the empirical conversion factors, we have derived the local \NIIab\ LF, which can be well fitted using a Schechter (1976) function with $\phi^*=10^{-3.83\pm0.09}$ Mpc$^{-3}$, $\alpha=-1.55\pm0.06$ and $L^* = 10^{7.36\pm0.04}$ $L_\odot$. 

\item
As a practical application of the \NIIab\ LF, we have also computed the SFR volume density for the local Universe using the newly derived SFR calibrator, which gives $\log\,\dot{\rho}_{\rm SFR} = −1.96\pm0.11$ $M_\odot$\,yr$^{-1}$\,Mpc$^{-3}$, in good agreement with the results of previous studies. 

\item
For a sub-sample of 12 LIRGs which have both \NIIab\ and \NIIbc\ lines observed, we have determined the electron density of the low-density ionized medium in these galaxies by matching the observed \NIIbc/\NIIab\ raitos ($R_{122/205}$) with theoretical grids, and find that $n_e$ is from $ < \sim 1$ cm$^{-3}$ to $\sim$66 cm$^{-3}$, with a median value of 22 cm$^{-3}$.
\end{enumerate}

\begin{acknowledgements}
We thank an anonymous referee for her/his careful reading and constructive comments. We thank Dr \'{A}lvaro Orsi for providing us his simulation results of the \NIIab\ LFs. Support for this work was provided in part by NASA through an award issued by JPL/Caltech. Y.Z. and Y.G. are partially supported by the National Natural Science Foundation of China under grants No. 11173059, 11390373 and 11420101002, and the CAS pilot-b project \#XDB09000000. VC would like to acknowledge partial support from the EU FP7 Grant PIRSES-GA-2012-316788. This research has made use of the NASA/IPAC Extragalactic Database (NED), which is operated by the Jet Propulsion Laboratory, California Institute of Technology, under contract with the National Aeronautics and Space Administration.
\end{acknowledgements}

%\end{CJK*}
\end{document}